\documentclass[preprint,superscriptaddress,notitlepage,float]{revtex4-2}
\usepackage{graphicx,bm,mathtools,color}
\usepackage[pdfstartview=FitH, CJKbookmarks=true, bookmarksnumbered=true, bookmarksopen=true, colorlinks, pdfborder=001, linkcolor=magenta, anchorcolor=blue, citecolor=magenta]{hyperref}
\usepackage{amsmath,amsfonts}
\usepackage{slashed}
\graphicspath{{Pictures/}}
\usepackage{lineno}
\usepackage{setspace}
\usepackage{comment}
\usepackage{titlesec}

\titleformat{\subsection}
  {\bfseries}     % heading style
  {}              % no number
  {0pt}           % no separation after number
  {}              % code before title

\titlespacing*{\subsection}
  {0pt}            % left indentation
  {\baselineskip}  % space before heading
  {0pt}            % space after heading

\newcommand{\bx}{\ensuremath{\mathbf{x}}}
\newcommand{\bq}{\ensuremath{\mathbf{q}}}
\newcommand{\dg}{^{\dagger}}
\newcommand{\vk}{\ensuremath{\mathbf{k}}}

\newcommand{\vq}{\ensuremath{{\mathbf{Q}}%_{\rm{afm}}
}}

\newcommand{\sgn}{{\rm sgn}}
\newcommand{\smb}{\ensuremath{\mathrm{SmB_6}}\ }
\newcommand{\blue}[1] {{\color{black}#1}}
\newcommand{\red}[1] {{\color{black}#1}}
\newcommand{\PC}[1]{{\color{black} #1}}
\begin{document} 
\renewcommand{\abstractname}{} 
%------------------

\title{Axionic tunneling from a topological Kondo insulator}

%---------------------------------------------
\author{Saikat Banerjee}
\affiliation{ Center for Materials Theory, Rutgers University, Piscataway, New Jersey, 08854, USA.}
\affiliation{Institute of Physics, University of Greifswald, Felix-Hausdorff-Stra{\ss}e 6, 17489 Greifswald, Germany.}
\author{Anuva Aishwarya}
\affiliation{Department of Physics, Harvard University, Cambridge, Massachusetts 02138, USA.}
\author{Fei Liu}
\affiliation{ State Key Laboratory of Optoelectronic Materials and Technologies,
Guangdong Province Key Laboratory of Display Material and Technology,China.} 
\affiliation{
School of Electronics and Information Technology, Sun Yat-sen University, Guangzhou 510275, China. }
\author{Lin Jiao}
\affiliation{Center for Correlated Matter and School of Physics, and  Zhejiang University, Hangzhou 310058, China. }
\author{Vidya Madhavan}
\email{vm1@illinois.edu}
\affiliation{ Department of Physics and Materials Research Laboratory, University of Illinois Urbana-Champaign, Urbana, IL 61801, USA.}
\author{Eugene J. Mele}
\email{mele@sas.upenn.edu}
\affiliation{Department of Physics, University of Pennsylvania, Philadelphia 19104, USA.}
\author{Piers Coleman}
\email{piers.coleman@rhul.ac.uk}
\affiliation{Hubbard Theory Consortium, Department of Physics, Royal Holloway, University of London, Egham, Surrey TW20 0EX, UK.}
\affiliation{ Center for Materials Theory, Rutgers University, Piscataway, New Jersey, 08854, USA.}
%---------------------------------------------
\date{\today}

\maketitle
%-------------------------------------

%-------------------------------------
\noindent{\bf \Large Abstract}

\noindent\hrulefill

%\linenumbers

\textbf{Discoveries over the past two decades have revealed the remarkable ability of quantum materials to emulate relativistic properties of the vacuum, from Dirac cones in graphene to \blue{Dirac} surface states of topological insulators. \blue{Yet one of the most elusive consequences of topology in quantum matter -- the axionic ${\bf E}\cdot{\bf B}$ contribution to the electromagnetic response, predicted to produce strong magnetoelectric effects -- remains experimentally challenging to detect.} Here we report \blue{evidence for an axion-like magnetoelectric response} obtained through scanning tunneling microscopy (STM) \blue{using a SmB$_6$ nanowire tip on an antiferromagnetic Fe$_{1+x}$Te sample}. Our measurements reveal a striking voltage-induced magnetization: millivolt biases generate \blue{measurable tip magnetizations} that reverse with the voltage. The magnitude, tunability, and reversibility of this signal are consistent with an axion-like \blue{${\bf E} \cdot {\bf B}$} coupling, \blue{which naturally accounts for the voltage-odd magnetic component of the tip spectral function while strongly constraining a conventional static-magnetism interpretation.} \blue{Moreover,} millivolt-scale control of spin polarization in a tunnel junction provides a new route for probing axionic electrodynamics and opens avenues for future STM and spintronic applications.} \\

\noindent{\bf \Large Main}

\noindent\hrulefill

The connection between bulk topology and boundary surface states reflects a deep universality between the anomalies of relativistic quantum mechanics and their solid-state counterparts in topological insulators. Topological states of matter represent a new frontier in the exploration of quantum materials with the potential for applications in spintronics and quantum information. An important platform for these phenomena is the class of topological Kondo insulators~\cite{Dzero2010,Dzero2016}, in which strong interactions between electrons in the bulk cause the boundary surface states to float at the Fermi energy, avoiding the need to fine-tune the chemical potential. The material SmB$_6$ has been of particular interest in this respect: the anticipated topological character of this insulator~\cite{Dzero2010,takimoto2011} is supported by the observation of  robust surface conductivity~\cite{wolgasttki,kimtki}, helical surface states detected in angle-resolved photoemission spectroscopy~\cite{Xu2014,Ohtsubo18}, reflectionless Klein tunneling through a potential barrier~\cite{Galitski2019} and the planar transmission of spin currents from a ferromagnet into the topological surface states~\cite{Liu2018}. 

The foundations of topology in quantum matter derive from mathematical index theorems~\cite{Atiyah_1975}, which predict \blue{that} the topology of an insulator \blue{can} manifest as a magneto-electric coupling \cite{Wilczek87,Zhang2008,Zhang_2009,Witten_2016} between electric (${\bf E}$) and magnetic (${\bf B}$) fields
%-----------------------------
\begin{equation}\label{eq.1}
\blue{S_\theta} =  \frac{e^2}{2\pi h}\int dt d\bx \, \theta(\bx) \, {\bf E} \cdot {\bf B},
\end{equation}
%-----------------------------
where \blue{\(S_{\theta}\) denotes the axion contribution to the electromagnetic action, and} $\theta(\bx)= \pm \pi$ acquires a non-zero value in a topological insulator ($e$ is the charge of the electron and $h$ Planck's constant). Theory predicts that these couplings are hidden unless there is broken time-reversal symmetry, in which case they manifest as a surface Hall conductivity  $\sigma_{xy}=\pm \frac{e^2}{2h}$~\cite{chang2023colloquium, xiao2018realization}. A second \blue{manifestation, which has been less explored in local tunneling probes, is the electric-field-induced magnetization associated with the axion magnetoelectric response}~\cite{Zhang_2009}. The axion coupling fixes the total induced magnetization, $\blue{{\bf M}_{\theta} = {\bf M}_{\rm orb}+{\bf M}_{\rm s} = \frac{\theta e^2}{2\pi h}\,{\bf E},}$ \blue{where ${\bf M}_{\rm orb}$ and ${\bf M}_{\rm s}$ denote, respectively, the orbital and spin contributions to the total magnetic response. The axion term constrains their sum, while their relative weights depend on the microscopic electronic structure.} Traditionally, Hall-current measurements have been used to probe the surface currents associated with the orbital magnetoelectric response ${\bf M}_{\rm orb}$~\cite{Mogi2017AxionInsulator,Zhang2023NonlocalAxion}. \blue{Here we show that magnetic-contrast tunneling can instead probe the spin-dependent component of the local tunneling spectral function, providing access to the local spin contribution ${\bf M}_{\rm s}$ induced by the electric field at the tunneling apex.} In particular, we present new experiments and analyses of magnetic tunneling into the topological insulator SmB$_6$ which \blue{provide evidence for this axion-like behavior and suggest that it can be probed within {\sl metallic} surface states.} In our study, we find that small \blue{($\sim 30$ mV)} voltages are sufficient to generate tip magnetizations of about 0.1$\mu_B$ which reverse with the bias voltage. We show that the atomic-scale tip to substrate distances generate strong electric fields \--- \blue{about $10^8-10^9$ V/m -- large enough to account} for the magnitude of the voltage-controlled magnetization.

\begin{figure}[htb!]
\centering
\includegraphics[width=0.8\textwidth]{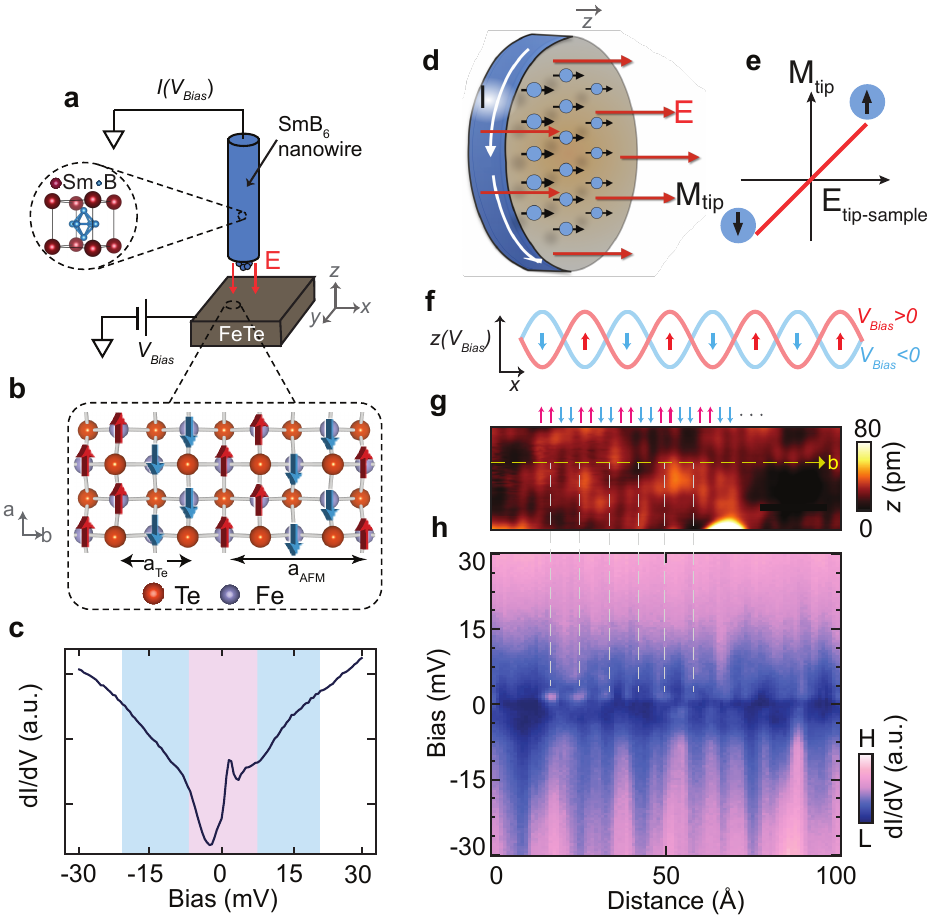}
\caption{\small \textbf{\small Scanning tunneling spectroscopy obtained using an SmB$_6$ nanowire tip on the lattice antiferromagnet  Fe$_{1+x}$Te.} {\bf a.} Schematic of the STM tunnel junction with an  SmB$_6$ nanowire  tip and  antiferromagnetic Fe$_{1+x}$Te  sample. {\bf b.} Graphic showing the bicollinear antiferromagnetic structure on the surface of Fe$_{1+x}$Te. \blue{The Fe moments form a canted magnetic structure with both in-plane and surface-normal components.} {\bf c.} Average $dI/dV$ spectra obtained with the nanowire tip on Fe$_{1+x}$Te. The blue shaded region highlights the Fano lineshape, and the pink shaded area within it denotes the feature associated with the topological surface state~\cite{Jiao2016}. {\bf d.} Electric field penetrating the  nanowire induces an axionic Hall current around the wire, and a spin magnetization at its tip. \blue{{\bf e.} The induced tip magnetization is linear in the applied electric field.} {\bf f.} \blue{The final atoms in the STM tip are} spin polarized by the axionic magnetization  leading to a voltage-tuned spin contrast. {\bf g.} Topography obtained with the SmB$_6$ tip on the surface of Fe$_{1+x}$Te at $T = 1.7$ K showing the spin contrast associated with the antiferromagnetic lattice. \blue{The crystallographic directions are indicated in the panel; the yellow dashed line is taken along the $\bf b$ direction and crosses the antiferromagnetic stripes shown schematically in panel {\bf b}.} The white dashed line marks the line along which point spectra shown in {\bf h.} have been obtained. {\bf h.} Differential conductance data ($\tfrac{dI(\bx)}{dV}$) collected over a 100\r{A} region \blue{along the same line cut} crossing several periods of the antiferromagnetic lattice constant ($T = 1.7$ K, $I_{\rm{set}}$ = 120 pA, $V_{\rm{Bias}}$ = 50 mV, $V_{\rm{mod}}$ = 600 $\mu $V). \blue{White vertical dashed lines serve as a guide to the eye, marking the phase of the magnetic contrast modulation along the spectroscopy line cut.}
}\label{fig:Fig1}
\end{figure}

Our tunneling experiments with an $\smb$ nanowire tip previously demonstrated magnetic contrast on an antiferromagnetic (AFM) substrate, Fe$_{1+x}$Te~\cite{Aishwarya2022}. \blue{The Fe moments in Fe$_{1+x}$Te form a canted antiferromagnetic
structure with both in-plane and surface-normal components~\cite{Singh2017}. The magnetic tunneling contrast therefore probes a projection of the staggered moment, rather than a moment oriented entirely perpendicular or parallel to the surface.} In particular, topographic scans at fixed current revealed a modulation in the height of the tip that \blue{mapped} out the staggered magnetization of the substrate (see Fig.~\ref{fig:Fig1}). Unlike a conventional magnetic tip~\cite{Aishwarya2022}, this spin contrast reverses with bias voltage. This unusual finding was previously interpreted as a signature of spin-momentum locking of the electrons in the \blue{Dirac} surface states.

\blue{Recent observations of similar voltage-reversing spin contrast in related tunneling experiments -- using SmB$_6$ tips on an altermagnetic substrate~\cite{Yang2026} and using a different topological-insulator tip, (BiBr)$_4$~\cite{Wang2025}, on related magnetic substrates -- make it particularly important to establish the microscopic origin of this effect. Here we present new measurements and analysis showing that the voltage-reversed spin contrast is most naturally described by a voltage-induced magnetic spectral component of the SmB$_6$ tip. This reassessment is motivated by the local weak-tunneling geometry of the STM experiment: the spin contrast is measured with a sharp tunneling tip at atomic-scale spatial resolution, so that the relevant quantity is the local apex spectral function rather than a momentum-selective planar tunneling current.} \blue{In this limit, summation over momenta removes the direct memory of spin-momentum locking (see Supplementary Information). Moreover, because the measured conductance lies deep in the weak-tunneling regime, the spin-sensitive current is determined by the equilibrium local Green's function of the tip apex. Contributions from the different nanowire facets may renormalize this local spectral function, and may enter with unequal weights, but they cannot generate a magnetic component when the facet--apex system preserves time-reversal symmetry. A detailed multi-facet calculation establishing this result is presented in the Supplementary Information.}

%----------------------------------------------------------------------------
%----------------------------------------------------------------------------
\clearpage
\pagebreak
\vfill\eject
 
%-----------------------------------------
\noindent{\bf  Origins of Spin Contrast}
%-----------------------------------------

\noindent\hrulefill

To %re-evaluate
understand the origins of the spin contrast, we have carried out a set of atomically resolved measurements of the conductance $\tfrac{dI(\bx)}{dV}= g(\bx,V)$, using an \smb nanowire \blue{tip} to reveal the spatially staggered, magnetic component of the local density of states of Fe$_{1+x}$Te. \blue{The local density of states $g(\bx,V)$ shown  in Fig.~\ref{fig:Fig1}c. displays the characteristic Fano line-shape of tunneling from the extended bulk SmB$_6$ within the tip.} We gathered high-resolution differential conductance data across multiple AFM periods of Fe$_{1+x}$Te, as illustrated in Fig.~\ref{fig:Fig1}h, from which we derived the Fourier transform $g({\bf q},V)$ of the position-dependent conductance, where $\bq$ is the wavevector. The Fourier transformed conductance $g(\vq,V) = \overline{\frac{dI(\bx_{\uparrow})}{dV} - \tfrac{dI(\bx_{\downarrow})}{dV}}$ at the magnetic wavevector $\vq$ describes the magnetic contrast between the conductance at sites $\bx_{\uparrow}$ and $\bx_{\downarrow}$ of the alternating up and down spins in the antiferromagnetic substrate, where the overbar indicates a spatial average.

\begin{figure}[htb!]
\centering
\includegraphics[width=1.0\textwidth]{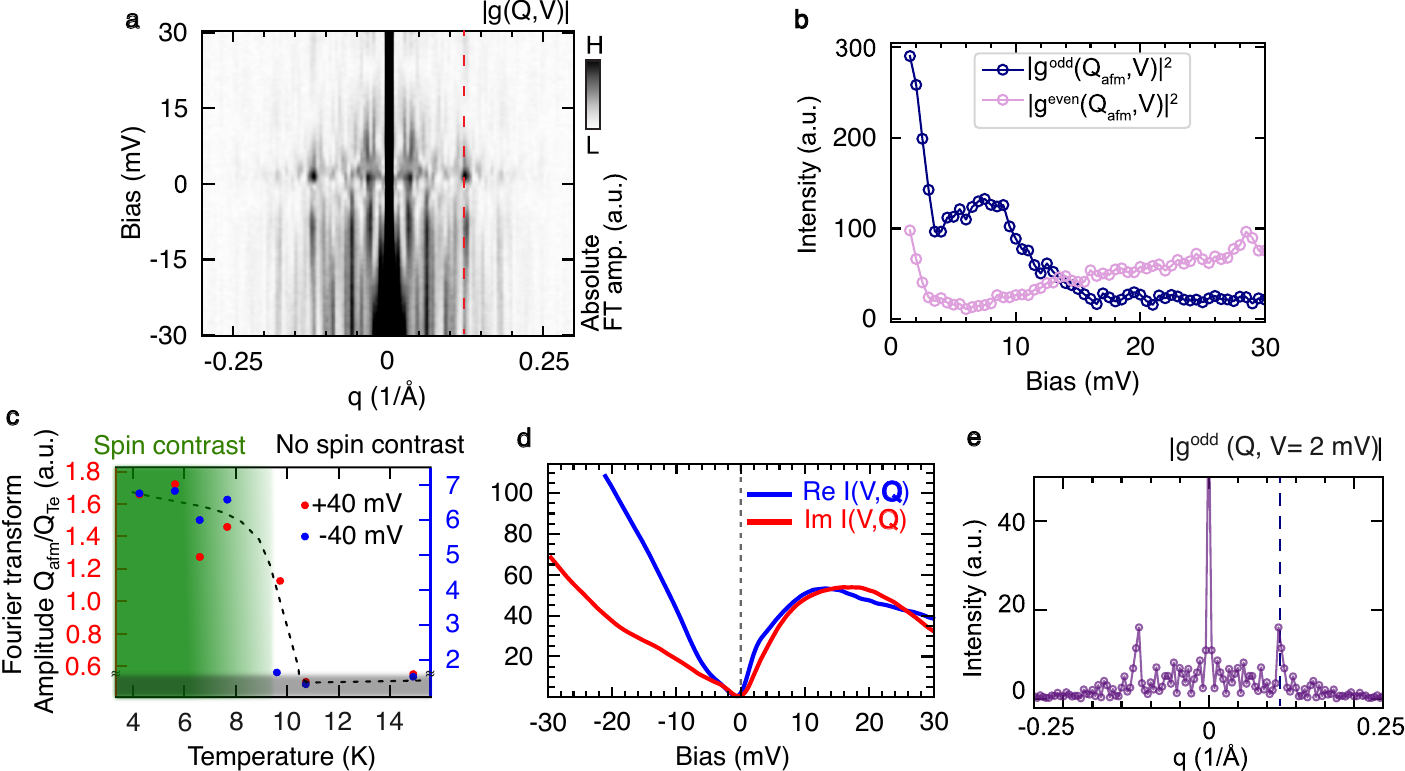}
\caption{\textbf{Odd-voltage signal  in the one-dimensional (1D) Fast Fourier Transform (FFT) of the $\tfrac{dI(\bx)}{dV}$ spectra at $\vq$.} \blue{The antiferromagnetic stripe period is $a_{\rm afm}\simeq 7.6~\text{\AA}$, corresponding to ${\bf Q}_{\rm afm}\simeq 0.13~\text{\AA}^{-1}$ in the Fourier-transform convention used here; the Te Bragg wave vector is ${\bf Q}_{\rm Te}\simeq 0.26~\text{\AA}^{-1}$.} {\bf a.} Absolute 1D FFT \blue{amplitude $|g({\bf q},V)|$} as a function of bias of the $\tfrac{dI(\bx)}{dV}$ spectra shown in Fig.~\ref{fig:Fig1}\blue{h}. \blue{The red dashed line marks ${\bf Q}_{\rm afm}$.} {\bf b.} Plot showing the relative contrast between the odd-voltage and even voltage signal at \blue{${\bf Q}_{\rm afm}$, corresponding to the red dashed line in panel a,} as a function of applied bias voltage. \blue{The vertical axis in panel b shows the Fourier intensity, proportional to the squared Fourier amplitude, in arbitrary units.} {\bf c.} Plot of the ratio of the intensity of the signal from ${\bf Q}_{\rm afm}$ to the Bragg peak (${\bf Q}_{\rm Te}$) as a function of temperature. Black dashed line is a guide to the eye. \blue{The spin contrast signal from the AFM stripes is highly suppressed near $\sim 10$ K, the proposed onset scale for the axion-like voltage-odd magnetic tunneling response.}
Reproduced from Ref.~\cite{Aishwarya2022}. {\bf d.} Interpolated (magnetic part) current-voltage plot obtained from the phase-referenced differential conductance data measured in our experiment. {\bf e.} Representative line cut of the odd-voltage signal at $V = 2$ mV as a function of wave-vector showing a clear signal at \blue{${\bf Q}_{\rm afm}$}.}\label{fig:Fig2}
\end{figure}

To characterize the dependence of this signal on voltage, we extracted the even and odd components of the voltage dependence, \blue{$g^{\rm even}({\bf Q},V)=\left[g({\bf Q},V)+g({\bf Q},-V)\right]$ and $g^{\rm odd}({\bf Q},V)=\left[g({\bf Q},V)-g({\bf Q},-V)\right]$.} \blue{By construction, $g^{\rm odd}({\bf Q},0)=0$ for an ideal voltage-symmetric measurement. The finite odd-voltage signal discussed below is therefore not a claim of a nonzero zero-bias odd component; it refers to the lowest finite-bias data points within the experimental voltage grid, with finite thermal and lock-in broadening.} Fig.~\ref{fig:Fig2}b compares these components within the low-energy range $V \in [0,5]$ \blue{mV}, \blue{showing that, away from strictly zero bias, the voltage-odd component is larger than the voltage-even component over the measured finite-bias range.} By contrast, the equivalent data for \blue{a} magnetic chromium tip is even in voltage [See extended figure~\ref{fig:Fig4}].

\blue{Since the magnetic contrast in the tunneling is governed by the spin-dependent local density of states of both the tip and the substrate, its leading magnetic contribution is proportional to the overlap between the tip magnetic spectral component and the staggered magnetic component of the substrate. In the low-energy limit used here, this reduces to $g(\mathbf Q,V)\propto {\bf m}_{\rm t}(V)\cdot{\bf m}_{\mathbf Q}$.} Our results therefore indicate a local \blue{magnetic spectral component ${\bf m}_{\rm t}(V)$} in the STM tip that reverses with voltage, \blue{${\bf m}_{\rm t}(V)\simeq-{\bf m}_{\rm t}(-V)$, rather than a purely static tip magnetization.}

%----------------------------------------------------------------------------
% Fig 2 -- Explaining the main reassessment of the experiment in SmB6
%----------------------------------------------------------------------------
%----------------------------------------------------------------------------
% Fig 2 -- Explaining the main reassessment of the experiment in SmB6
%----------------------------------------------------------------------------
\clearpage

\blue{A substrate-only gating or band-structure effect may modify the scalar local density of states of Fe$_{1+x}$Te, but it does not account for the key combination of observations. The measured signal is the magnetic Fourier component at the AFM wave vector, and its voltage reversal is observed with SmB$_6$ tips but not with conventional Cr tips on the same substrate. The data therefore require a spin-dependent, voltage-odd component of the SmB$_6$-controlled tunneling channel, rather than a purely electrostatic band-structure reconstruction of the Fe$_{1+x}$Te substrate.}

To obtain an estimate of the magnitude of the voltage-induced \blue{magnetic spectral component of the SmB$_6$ tip}, we compared the spin contrast with a conventional chromium magnetic tip tunneling into the same Fe$_{1+x}$Te substrate. We define the normalized spin contrast $\Psi(V)$ at a fixed bias voltage $V$, as  the ratio $\Psi(V)= \tfrac{ I(V,\vq)}{I(V,{\bf Q}_{\rm Te})}$ between the \blue{Fourier component of the} tunneling current at the magnetic wavevector  $\vq$ and the \blue{Fourier component at the} wavevector ${\bf Q}_{\rm Te}$ of the non-magnetic tellurium atoms. The comparison of this normalized ratio for Cr and \smb tips, provides \blue{an order-of-magnitude estimate of} the ratio of \blue{effective tip magnetic} moment sizes, 
%-----------------------------
\begin{equation}\label{eq.2}
\frac{\Psi_{\rm Sm}(V)}{\Psi_{\rm Cr}(V)} \blue{\simeq} \frac{{\rm m_{\rm Sm}}{\PC {(V)}}}{\rm m_{\rm Cr}}.
\end{equation}
%-----------------------------
By comparing this ratio for the Cr and \smb $\,$ tips we find that $\left\vert \tfrac{m_{\rm Sm}}{m_{\rm Cr}} \right\vert \sim 0.13$ at a bias voltage of $\sim 35$ \blue{mV} (see \hyperref[met:Moment]{Methods}). Bulk Cr has a staggered magnetization of 3.6$\mu_B$, so our results indicate a tip magnetization of order $0.4 \mu_B$ per Sm atom at this voltage. \PC{The voltage reversal of the spin contrast therefore indicates a voltage-controlled magnetoelectric response of the \smb tip. Moreover, since magnetization and electric field have opposite parities under time reversal and inversion, the reversal of the magnetization with voltage, $m_{\rm {Sm}}{(V,T)} \sim \Psi(V,T)  {\sgn}\PC{V}$ implies the development of a spin-contrast order parameter $\Psi(V,T)$ that is odd under time-reversal and under inversion.}

\blue{The temperature dependence of the voltage-reversed spin contrast, $\Psi(V,T)$, is shown in Fig.~\ref{fig:Fig2}c. The signal is strongly suppressed above $\sim 10\,{\rm K}$, far below the N{\'e}el temperature ($T_N \approx 60-70$ K) of Fe$_{1+x}$Te~\cite{Enayat2014,Singh2017}. This temperature scale is therefore not associated with the loss of magnetic order in the Fe$_{1+x}$Te substrate. This conclusion is further supported by the Cr-tip control measurement: a conventional magnetic Cr tip tunneling into the same Fe$_{1+x}$Te substrate images the antiferromagnetic order without an analogous voltage reversal~\cite{Aishwarya2022}.} \blue{The disappearance of the voltage-reversed contrast near
$\sim 10$~K therefore identifies a characteristic temperature scale of the SmB$_6$ tip, or of the SmB$_6$-controlled tunneling channel, rather than of the Fe$_{1+x}$Te substrate. Together with the Cr-tip control, this strongly constrains explanations based on substrate-only gating, band-structure changes, or surface reconstruction.} \blue{From this comparison, we conclude that the voltage-induced magnetic spectral component is associated with \smb, signaling the development of a time-reversal-odd \PC{and inversion-odd} response below $10$ K.} The absence of any bulk anomalies in the specific heat indicates that this is not a bulk phenomenon, \PC{suggesting the broken symmetry resides on the surface}. 

Rather general topological arguments~\cite{Zhang2008,Vanderbilt2009,fijalkowski2021any} tell us that if broken time-reversal symmetry develops at the surface below 10 K, a topological insulator can develop a magneto-electric polarization ${\bf M} = \frac{\theta}{\pi} \frac{e^2}{2h} \,{\bf E}$. To obtain an order-of-magnitude estimate of the corresponding local magnetic moment per Sm atom, we write $(m_{\rm Sm}/\mu_B) = M a^3/\mu_B$, where $a^3$ is the unit cell volume of SmB$_6$. This gives
%-------------------------------------
\begin{equation}\label{eq.3}
\frac{m_{\rm Sm}}{\mu_B} = \left(\frac{a}{a_B}\right)^3\frac{E}{2 \pi E_H},
\end{equation}
%-------------------------------------
where $a_B = \frac{\hbar}{m c \alpha} \approx 5.2\times 10^{-11}$ m and $E_H = \frac{e}{4 \pi \epsilon_0 a_B^2}\approx 0.51 \times 10^{12}$ ${\rm V}/{\rm m}$ are the Bohr radius and the associated atomic electric field. 

\begin{figure}[t] 
\centering
\includegraphics[width=\textwidth]{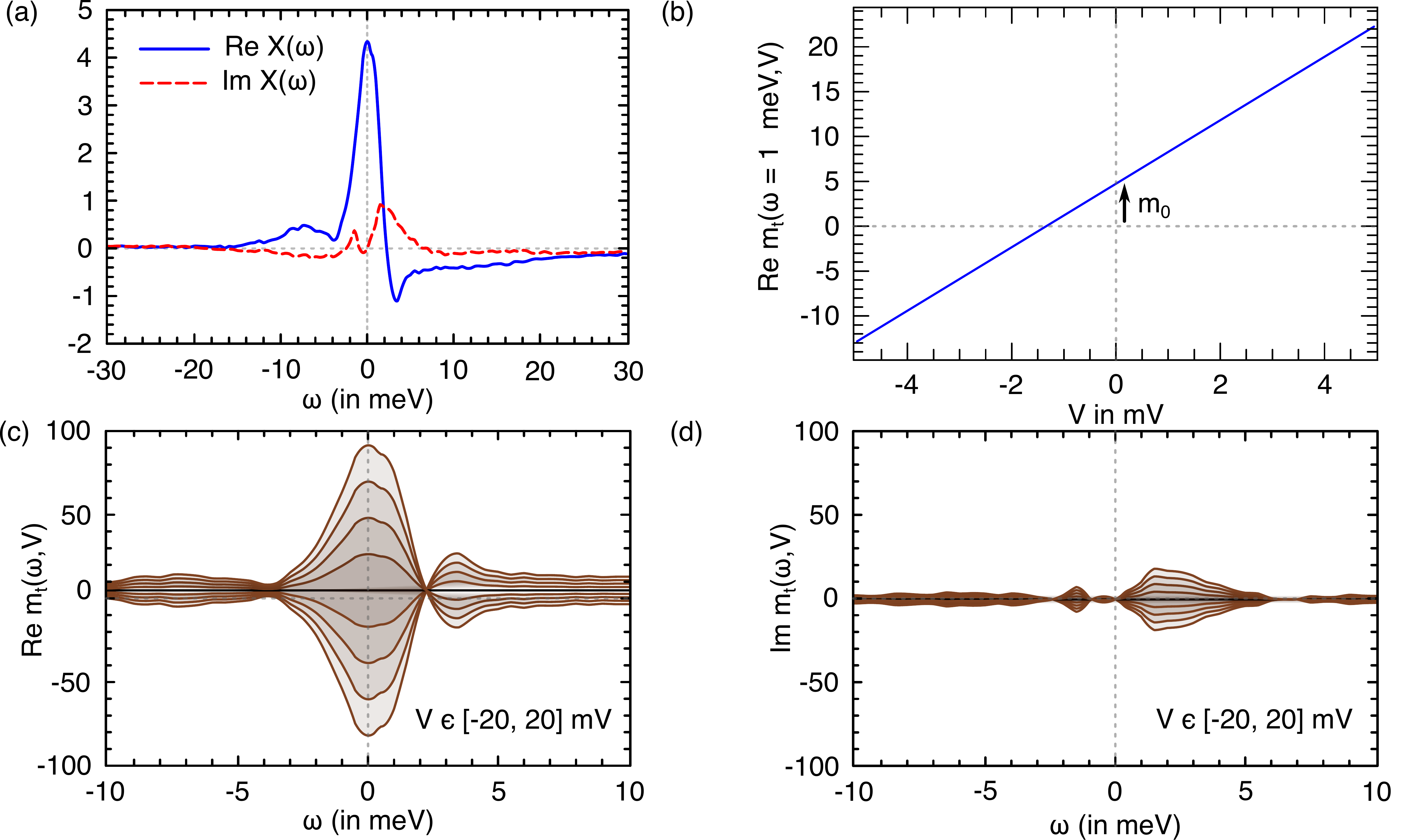}
\caption{\blue{\textbf{Voltage-induced magnetic response spectrum.} \textbf{a}. Real and imaginary parts of the extracted voltage-induced magnetic response function $X(\omega)$. \textbf{b}. Real part of the reconstructed magnetic signal 
$m_{\rm t}(\omega=1\,{\rm meV},V)=V X(\omega=1\,{\rm meV})+m_0$ as function of applied bias voltage. The small zero-bias intercepts indicate the corresponding real part of the residual component $m_0$. \textbf{c},\textbf{d}. Real and imaginary parts, respectively, of the reconstructed voltage-tuned magnetic spectrum  $m_{\rm t}(\omega,V)=V X(\omega)+m_0$ for applied bias voltages $V=-20,-15,\ldots,20\,{\rm mV}$, in increments of $5\,{\rm mV}$. All quantities are plotted in arbitrary units using the same normalization as the phase-referenced differential conductance data.}}\label{fig:Fig3}
\end{figure}

\blue{In the STM geometry the relevant electric field is the local field at the SmB$_6$ apex/surface region, which we estimate as $E\sim V/d_{\rm eff}$, where $d_{\rm eff}$ is the effective length scale over which the voltage drops near the tunneling apex. For representative tunneling parameters, $V\sim 40$ mV and $d_{\rm eff}$ on a sub-\AA ngstr\"om to \AA ngstr\"om scale give a local field of order $E\sim 10^8$--$10^9$ V/m. This corresponds to $E/E_H\sim 10^{-4}$--$10^{-3}$, well within the linear-response regime. Using $a=4.1\times 10^{-10}$ m for SmB$_6$, Eq.~\ref{eq.3} gives a moment scale $m_{\rm Sm}\sim 10^{-2}$--$10^{-1}\mu_B$ per Sm atom, reaching $m_{\rm Sm}\sim 0.15\mu_B$ for the largest local fields. This estimate should be regarded as an order-of-magnitude consistency check rather than a fit: it shows that realistic STM fields are large enough to generate a detectable local magnetic spectral component, while remaining far below the atomic saturation scale. The sign of this contribution follows the sign of the applied voltage, naturally accounting for the observed voltage reversal, cf. Fig.~\ref{fig:Fig3}a.} 
\vfill\eject
%--------------------------------------------------------
\noindent{\bf Extracting the Surface Magnetic Spectrum}
%--------------------------------------------------------

\noindent\hrulefill

Our tunneling results allow us to extract the magnetic spectrum of our \smb tunneling tip. The tunnel current  is given by the Bardeen formula (see \hyperref[seq.5]{Methods})
%------------------------------
\begin{equation}\label{eq.4}
I( V,{\bf x}) = \frac{2\pi e|t|^2}{\hbar}\int d\omega {\rm Tr}\left[
g_{\rm{t}} (\omega)g_{\rm{s}} (\omega-eV,{\bf x} ) \right]
\big[f(\omega-eV)-f(\omega)\big],
\end{equation}
%------------------------------
where $g_{\rm{s}} (\omega,\bx ) = \frac{1}{\pi}{\rm Im}{\cal G}_{\rm{s}} (\omega-i\delta,\bx ) $ and $g_{\rm{t}} (\omega ) = \frac{1}{\pi}{\rm Im}{\cal G}_{\rm{t}} (\omega-i\delta ) $ are the energy-resolved spin-density matrices of the substrate (s) and tip (t), while  $f(\omega) = 1/(e^{\beta\omega}+1) $ is the Fermi-Dirac function. \blue{Following Ref.~\onlinecite{PhysRevLett.103.206402}, here we assume $g_{\rm{t}} (\omega)$ is the spectral function of an effective tunneling field combining conduction- and composite $f$-electron channels. It therefore already incorporates cotunneling and the associated interference responsible for the Fano line shape, with the Kondo material here forming the tip rather than the substrate.} If we Fourier transform over the tip position, replacing the tip co-ordinate $\bx$ by the wavevector of the Fourier transform ${\bf q}$, we obtain
%------------------------------
\begin{equation}\label{eq.5}
I( V,{\bf q}) = \frac{2\pi e|t|^2}{\hbar}\int d\omega {\rm Tr}\left[
g_{\rm{t}} (\omega)g_{\rm{s}} (\omega-eV,{\bf q} ) \right]
\big[f(\omega-eV)-f(\omega)\big].
\end{equation}
%------------------------------
\blue{At the AFM wave vector $\bq=\vq$, the substrate Green's function contains the staggered magnetic component. Since the Fe moments are canted, we retain the surface-normal projection probed in the present tunneling geometry, $g_{\rm s}(\vq,\omega) = m_{\rm s}\sigma_z$, where $m_{\rm s}$ denotes the surface-normal component of the staggered moment, rather than the full Fe moment. At the low energies of our experiment, we neglect the energy dependence of $m_{\rm s}$.} If we decompose  the tip density of states into a paramagnetic and magnetic part, $g_{\rm t} = \rho_{\rm t}(\omega) + \tilde m_{\rm t}(\omega,E ) \sigma_z$, where $E = V/d_{\rm eff}$ is the strength of the surface electric field for a tip to substrate separation $d_{\rm eff}$, the tunnel current at the magnetic $\vq$ vector is then 
%------------------------------
\begin{equation}\label{eq.6}
I (V,\vq) 
= I_0\int d\omega 
\big[f(\omega-eV)-f(\omega)\big]m_{\rm{t}}(\omega,V) ,
\end{equation}
%------------------------------
where $m_{\rm t}(\omega,V)= \tilde m_{\rm t}(\omega,E)\vert_{E=V/d_{\rm eff}}$ is the magnetic density of states in the tip induced by the tunneling field  and $I_0=4\pi e|t|^2 m_{\rm s}/{\hbar}$. Fig.~\ref{fig:Fig2}(d) shows the magnetic component of the tunneling current. \PC{Unlike conventional magnetic tunneling, the current $I (V,\vq) $   {\sl does not} reverse with voltage, reflecting the reversal of
the differential conductance with voltage. }

\blue{Here, we describe the magnetic spectral density of the tip phenomenologically as}
%------------------------------
\begin{equation}
\blue{
m_{\rm t}(\omega,V)=V X(\omega)+m_0,
}
\end{equation}
%------------------------------
\blue{where $X(\omega)$ is the voltage-induced magnetic response function and $m_0$ represents a voltage-independent residual magnetism.} \PC{Introducing the magnetic conductance ${\sf G}(eV) = {I(V,\vq)}/{(I_0 e V)}$, we then obtain $m_0= {\sf G}(0)$ and 
%------------------------------
\begin{equation}
{\sf G}'(eV) =  \int_{-\infty}^{\infty} \bigl(- f'(\omega - eV) \bigr)X(\omega) d\omega
\end{equation}
%------------------------------
where primes denote differentiation (see~\hyperref[met:Spectrum_Analysis]{Methods}). The function $-f'(\omega)$ is a thermally broadened delta function, and at low temperatures, the lack of sharp features in the magnetic conductance ${\sf G}(eV)$ then allows us to replace $X(eV) ={\sf G}'(eV)$.}
\blue{In practice, we extract $G(eV)$ from a smooth complex interpolation of the phase-referenced magnetic conductance. This gives a stable reconstruction of both the real and imaginary parts of $X(\omega)$ and separates the voltage-induced response from any residual magnetization $m_0$.}

\blue{Fig.~\ref{fig:Fig3}(a) shows the real and imaginary parts of the extracted voltage-induced magnetic response function $X(\omega)$. Fig.~\ref{fig:Fig3}(b) shows the real and imaginary parts of the reconstructed magnetic signal $m_{\rm t}(\omega=1\,{\rm meV},V)$ as functions of applied bias voltage, with the zero-bias intercepts identifying the finite real part of the residual component $m_0$. Fig.~\ref{fig:Fig3}(c,d) show the real and imaginary parts, respectively, of the reconstructed voltage-tuned magnetic signal $m_{\rm t}(\omega,V)=V X(\omega)+m_0$ for applied bias voltages $V\in[-20,20]\,{\rm mV}$. The residual component at this energy is finite but small, indicating that the low-energy magnetic spectral weight of the \smb tip is dominated by the voltage-induced response rather than by a conventional static magnetic polarization.} %The phase of the tunnel current $I(\omega,\mathbf{Q})$ has been referenced to its value at zero bias. 
\blue{The complex magnetic conductance is phase-referenced to its value at $\omega=0$.  The extracted response function $X(\omega)$ is predominantly real. At $\omega=1\,{\rm meV}$, and for bias voltages $|V|\leq 2\,{\rm mV}$ corresponding to energies $|eV|\leq 2\,{\rm meV}$ within the bulk gap, the reconstructed magnetic signal is dominated by the voltage-induced response.}

\blue{The dominance of the voltage-linear contribution, $m_t(\omega,V)\simeq V X(\omega)$, shows that the magnetic spectral weight of the \smb tip is primarily generated by the applied electric field, consistent with an axion-like magnetoelectric response. A notable feature of the extracted spectrum is that the real part of $X(\omega)$ changes sign near $\omega\simeq 2.3$ meV. This sign reversal corresponds to a $\pi$ phase shift of the magnetic Fourier component, or equivalently to an interchange of the apparent spin contrast at higher energies. Since this energy lies in the range where the non-magnetic tunneling spectrum shows Fano-like structure, the sign change may be related to interference between tunneling through Sm $f$- and $d$-derived states~\cite{Hoffmann23,JP2013}.} \\

%----------------------------------------------------------
% Fig 3 -- Our interpretation for the axionic explanation
%----------------------------------------------------------
%-----------------------------------------------
\clearpage

%----------------------------------------------------------------
\noindent{\bf  Axionic behavior with Metallic Surface States}
%----------------------------------------------------------------

\noindent\hrulefill

\blue{The observation of axionic behavior in a topological insulator with gapless surface states is at first surprising. Axionic behavior has been detected in several insulating systems through Hall currents associated with an electric-field-induced orbital magnetization, for which the electric field must penetrate the material rather than be screened by mobile surface carriers~\cite{Mogi2017AxionInsulator,Zhang2023NonlocalAxion}. However, penetration of the normal field component will occur in the presence of  metallic surface states and the underlying topology and index theorems can accommodate metallic, gapless boundaries~\cite{Witten_2016,Bergman2011,Qi2011,Atiyah_1975}. Although time-reversal-symmetry breaking opens a gap at the Dirac point, angle-resolved photoemission shows that the extrapolated Dirac points of the three SmB$_6$ surface cones are buried in the valence or conduction bands~\cite{ARPES2013b,ARPES2013}. Weak time-reversal-symmetry breaking therefore need not gap the metallic surface states.

These metallic surface states screen the electric field from penetrating deeply into the bulk, but permit it to enter over their finite Thomas--Fermi screening length. The resulting near-surface field induces a voltage-sensitive magnetization ${\bf M}={\bf M}_{\rm orb}+{\bf M}_{\rm s} =\frac{\theta}{\pi}\frac{e^2}{2h}\,{\bf E}$, containing both orbital and spin components. The magnetoelectric response is therefore generated in an extended near-surface region of the SmB$_6$ apex, involving many atoms within the screening and surface-state penetration length. The outermost tunneling orbital does not itself constitute the axionic medium; rather, its spin-dependent local spectral function provides an atomic-scale readout of the polarization inherited from the surrounding SmB$_6$ region. The relevant electronic boundary region is bounded by the surface-state penetration depth, estimated to be less than approximately $30$~nm~\cite{Liu2018}. This is smaller than, or at most comparable to, the $30$--$50$~nm nanowire radius~\cite{Aishwarya2022}, and is distinct from the \AA ngstr\"om-scale vacuum tunneling distance, which controls the tunneling matrix element and spatial resolution.}

Axionic behavior does, however, require time-reversal symmetry breaking at the surface, for while axion fields of $\theta=\pm\pi$ in the bulk do not break time-reversal symmetry, intermediate values at the surface do so. \blue{The strong rise of the voltage-odd magnetic tunneling below approximately $10$~K suggests that the SmB$_6$ surface/apex region develops a time-reversal-odd response below this scale. Although there are many superficial observations consistent with this hypothesis, such as the sudden low-temperature onset surface-dominated transport ~\cite{wolgasttki,kimtki}, the development of spectroscopic and magnetic anomalies on a comparable $10$--$15$~K scale~\cite{Jiao2016,Neupane2013,PhysRevB.89.161107} and the report  of  low-temperature hysteretic or history-dependent behavior in surface-sensitive measurements~\cite{Nakajima2016,PhysRevB.92.115110}, definitive  tests of the surface axion hypothesis are needed.} \\

%------------------------------------
\begin{comment}
{\color{black} Do we need this paragraph? \red{No we do not need this, why should we speculate on the possible origin of TR breaking!}}
\blue{One possible microscopic origin of this surface time-reversal symmetry breaking is the modification of the local Kondo scale at the SmB$_6$ surface~\cite{PhysRevLett.114.177202}. Modified coordination and bonding can weaken the effective Sm $4f$--$5d$ hybridization, leaving partially unscreened Sm moments that interact through the metallic surface states via RKKY exchange~\cite{PhysRevLett.102.156603}. Depending on the surface termination and band structure, these interactions may favor
ferromagnetic or antiferromagnetic correlations. We emphasize that this is a possible mechanism for the required surface symmetry
breaking, rather than a conclusion of the present experiment.}

\blue{The present temperature-dependent data do not yet establish whether this onset is a thermodynamic phase transition or determine its universality class. We therefore regard $\theta$ phenomenologically as a collective near-surface field and leave the microscopic nature of the approximately $10$~K onset to future temperature-, field-, and bias-history-dependent measurements.} \\
\end{comment}
%------------------------------------

\blue{
%The present temperature-dependent data do not establish whether the suppression near $10$K represents a thermodynamic phase transition or determine its universality class. 

\noindent{\bf Outlook}

\noindent\hrulefill

Our interpretation of $\theta$ as a surface order parameter leads to sharp experimental predictions. If the low temperature onset of magneto-electric effects reflects axionic order, then the normal gradient of the axion field
becomes an emergent Ising order parameter:
\[
\sigma = \sgn \left( {\bm \nabla}\theta \cdot \hat{\bf n}\right)= \pm 1, \quad M_z \propto \sigma V,
\]
If this is the case, then cooling an SmB$_6$ tip through the onset will cause the sign of the spin contrast to randomly reset.  Moreover,  simultaneous electric- and magnetic-field cooling should train its sign according to $\sigma \propto {\bf E} \cdot {\bf B}$. By contrast, if the observed contrast originates from helical tunnelling, its sign is fixed by the tip geometry and will remain unchanged across repeated thermal cycles. Temperature-, field-, and bias-history-dependent experiments can therefore distinguish the two mechanisms.}

{%\color{blue}
%----------------------------------------------------------
\begin{figure}[t] 
\centering
\includegraphics[width=0.6\textwidth]{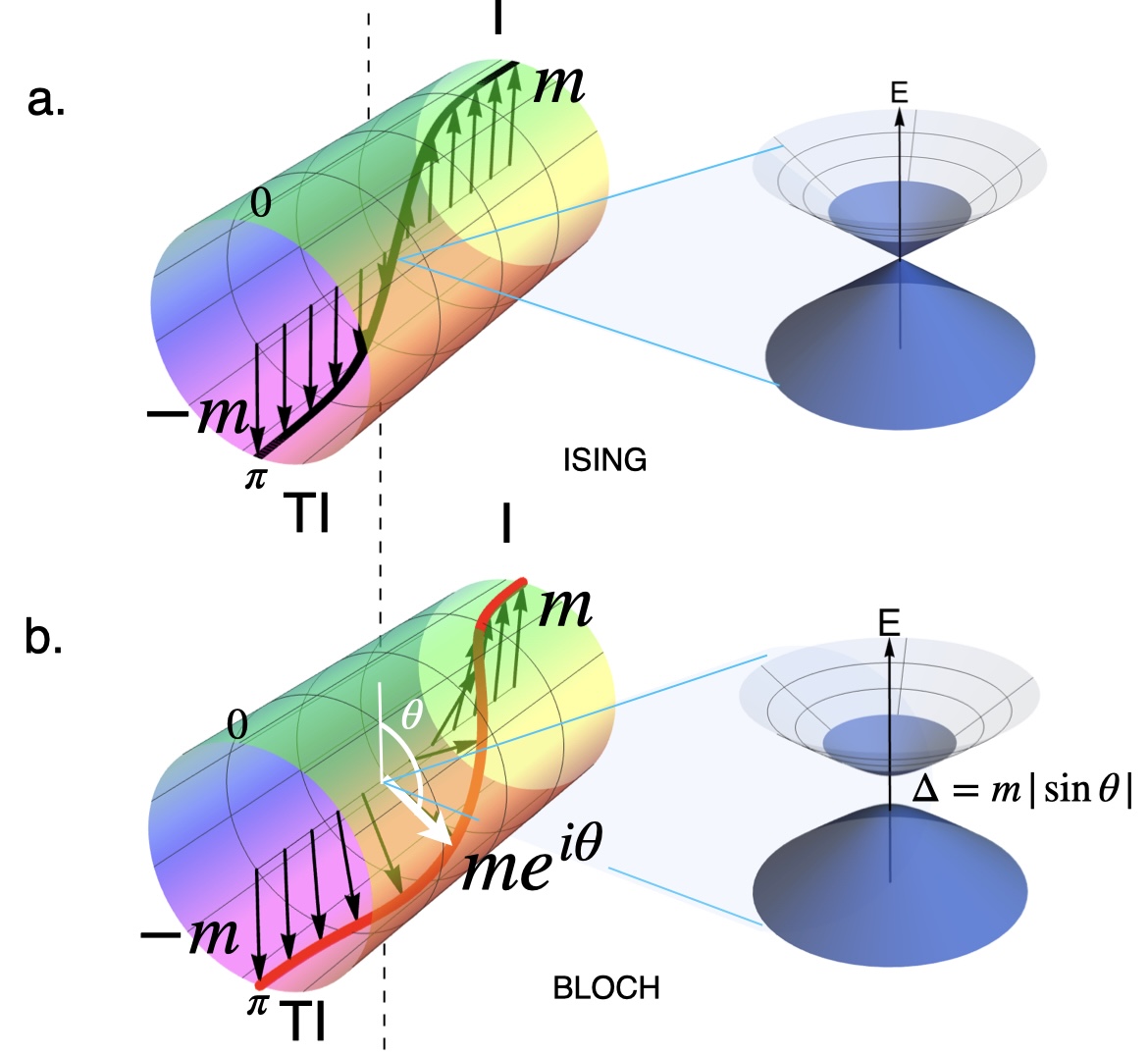}
\caption{\blue{\textbf{Scenario for a topological domain wall} between a bulk topological insulator (TI) and a conventional insulator (I). \textbf{a}. For weak interactions, the surface of a topological insulator forms an Ising domain wall, in which the Dirac mass passes through zero at the surface, giving rise to gapless surface states with a conical dispersion. \textbf{b}. For strong interactions, a Bloch domain wall develops in which the Dirac mass acquires an imaginary part, giving rise to axionic properties. At the surface, the node of the surface states acquires a gap that is buried beneath the Fermi surface, thus preserving metallic surface properties.}}\label{fig:Fig4X}
\end{figure}
%-----------------------------------------------

\blue{The  axionic interpretation of SmB$_6$ suggested by our experiments leads to a new way of thinking about the surfaces of interacting topological insulators.  We can regard the surface of a topological insulator as an interface between a topological and conventional insulator.  Near the gap-inverting quantum phase transition from  a conventional insulator (I) to a topological insulator(TI), the low-energy physics is governed by an emergent Dirac equation. In the conventional picture, the interface between the two insulators is a Jackiw-Rebbi soliton~\cite{JackiwRebbi76} at which the Dirac mass abruptly reverses, giving rise to protected gapless surface states (see Fig.~\ref{fig:Fig4X}a). 

Interactions between electrons introduce a richer set of possibilities, for the Dirac equation admits the possibility 
of a continuous rotation between positive and negative mass via a family of  axionic Hamiltonians $H(\theta) = {\bm \alpha} \cdot {\mathbf p} + \beta m_c(\theta)$ with a complex mass $m_c(\theta)=m e^{i\theta\gamma_5}$ where $\alpha $ and $\beta$ are the Dirac matrices, and  $\gamma_5=i \beta \alpha_1\alpha_2\alpha_3$ is the axial  matrix~\cite{Wilczek87}. From this perspective, the axion field $\theta(x)$ is the phase of a complex order parameter ${\cal M}(x)=m_e e^{i\theta(x)}$.

In a non-interacting topological insulator, the axion angle abruptly jumps  from $\theta=0$ to $\pm\pi$ at the surface forming an Ising domain wall. Interactions allow the complex order parameter ${\cal M}(x)$ to develop a phase stiffness, lowering the energy by forming a Bloch domain wall in which $\theta$ evolves smoothly from $\theta=0$  to $\pm\pi$ without the mass $|{\cal M}(x)|$ ever vanishing (Fig.~\ref{fig:Fig4X}b). A Ginzburg-Landau description of this physics~\cite{Bulaevskii1964DomainWall,Woodford2008Bloch} is
%-----------------------------
\begin{equation}
F = \frac12\int dx
\left\{
|\partial_x{\cal M}|^2
+\left(|{\cal M}|^2-1\right)^2
+\frac{\Delta(x)^2}{\lambda}
\right\},
\end{equation}
%-----------------------------
where ${\cal M}(x)=|{\cal M}(x)|e^{i\theta(x)}=m(x)+i\Delta(x)$ is the complex Dirac mass, $x$ is the coordinate normal to the surface, and $\lambda$ parametrizes the electron interactions that couple to the anomalous mass term. Deep inside the topological insulator, ${\cal M}(-\infty)=-1$, whereas outside, ${\cal M}(\infty)=+1$, so a domain wall interpolates between these limits.

For weak interactions $\lambda<1$, the stable solution is the familiar Ising wall, $m(x)=\tanh  x$ with $\Delta(x)=0$ (see Fig.~\ref{fig:Fig4X}a), but  for stronger interactions $\lambda > 1$, the minimum-energy configuration is a Bloch wall (see~\hyperref[eq:domains]{Methods}), where now
\[
{\cal M}(x) = \mathrm{tanh}(\frac{x}{\sqrt{\lambda}})  \pm i 
\sqrt{1-\frac1\lambda}\,
\mathrm{sech}(\frac{x}{\sqrt{\lambda}}),
\]
breaks time-reversal and inversion symmetry at the interface (see Fig.~\ref{fig:Fig4X}b).  

We note finally that lattice models of interacting topological surface states are slices of four-dimensional Chern insulators,  where the lattice momentum of  the fourth dimension is set to the axion angle $\theta$ \cite{Zhang2008}. This allows $\Delta $ to be tentatively identified as a hybridization between $d$ and $f$ orbitals at the surface Sm site (see \hyperref[eq:Wilson_Dirac]{Methods}). Normally, $d$ and $f$ orbitals do not mix inside a single atom, but the broken inversion symmetry at a surface allows this mixing to take place. The sign of $\Delta$ sets  the surface parity anomaly by selecting one of the two time-reversed Bloch-wall configurations, thereby fixing the sign of the associated Hall and magnetoelectric response. Kondo or other electron exchange interactions are the likely origin for these anomalous terms. As previously noted, even though the Dirac nodes of the surface states will develop a gap, in SmB$_6$ they are buried in the valence band, allowing the surface states to remain gapless even in the presence of an axionic domain wall.}

%-----------------------------
%{I moved this content to the second last paragraph in the previous section, seems to fit better! Please chnage back if you prefer!}
\begin{comment}
Our results can be tested in future experiments. The surface normal gradient of an axion angle, ${\bm \nabla} \theta \cdot \hat {\bm n}$, acts as an Ising order parameter, $\sigma=\pm1$ which sets the sign of the coupling between the  bias voltage and the tip magnetization. 
\[
M_z\propto \sigma V.
\]
Consequently, each time an %sample
SmB$_6$ tip is cooled through the putative surface transition temperature $T_c$, the order parameter should spontaneously choose either sign, leading to random reversals of the spin contrast between cool downs. Furthermore, field cooling in finite electric and magnetic fields will train the sign of axionic order according to $\sigma \propto {\bf E} \cdot {\bf B}$, thereby selecting a preferred magnetoelectric response. By contrast, if the observed spin contrast originates from helical tunneling, its sign will be fixed by the tip geometry and remain unchanged across repeated thermal cycles. This distinction provides a sharp experimental test of the proposed mechanism.
\end{comment}
%-----------------------------

\blue{Finally, if these predictions are confirmed experimentally, they would establish SmB$_6$ as an interacting axionic conductor. More generally, the ability to control tunnel-junction spin polarization using only millivolt-scale bias voltages points towards new ways of probing axionic electrodynamics and may provide a foundation for future STM-based probes and low-energy spintronic devices.}

\clearpage

\noindent \textbf{Methods} 

%--------------------------------------------------------
\subsection*{\textbf{Scanning Tunneling Measurements:}\label{met:STM}} 
%--------------------------------------------------------

\noindent The STM experiments were performed in a custom Unisoku STM (USM 1300) that can operate at 300 mK and above. Single crystals of FeTe were cleaved at ~90 K in UHV and immediately inserted into the STM. $dI/dV$ spectra were collected using a standard lock-in technique at a frequency of 893.4 Hz. FeTe crystals were studied at 1.7 K unless otherwise mentioned. The protocol for fabricating the nanowire tips is mentioned elsewhere~\cite{Aishwarya2022}. \blue{Because SmB$_6$ tips may have either Sm or B terminations, we do not assign a specific atomic identity to the terminal atom; the low-energy tunneling spectral weight nevertheless derives from Sm $d$- and $f$-states, either directly or through boron atoms. The tips occasionally pick up Fe atoms from the substrate, in which case the spin contrast no longer reverses with bias. Voltage reversal in both the topography and $dI/dV$ is restored after cleaning the tip with a bias pulse, allowing contaminated-tip data to be excluded. Conventional Cr tips do not exhibit voltage reversal. The spatially resolved $dI/dV$ line scans analysed in Fig.~\ref{fig:Fig1}h were newly acquired for the present work; previously published topographic data reproduced for comparison are identified explicitly in the corresponding captions.}

The experimental data were processed in Python using standard NumPy functions for the one-dimensional Fourier transforms, plotting, and visualization. The odd- and even- voltage Fast Fourier transformed (FFT) signal were calculated from the $dI/dV$ spectra or $g(V,\bx)$ using the formulae $g^{\rm{odd}}(V, \vq) = {\rm FFT}[g(V, \bx) - g(-V, \bx)]$ and $g^{\rm{even}}(V, \vq) =  {\rm FFT}[g(V, \bx) + g(-V, \bx)]$.  

\blue{To retain the relative phase of the modulations at opposite biases, we use the phase-referenced Fourier signal. Writing
$z(\bq,E)=|z(\bq,E)|e^{i\theta(\bq,E)}$, we define}
%------------------------------------
\begin{equation}
\label{eq.phase_ref}
\blue{z_{\rm PR}(\bq,|E|) = |z(\bq,E)| \cos\!\left[ \theta(\bq,E)-\theta(\bq,-E) \right].}
\end{equation}
%------------------------------------
\blue{This removes the arbitrary Fourier phase while preserving the relative phase between positive and negative bias. A negative $z_{\rm PR}$ at the magnetic wavevector indicates a reversal of the magnetic contrast. The same phase-referencing procedure is applied to the differential-conductance data.}

%-----------------------------------------------------------------
\subsection*{\textbf{Tunnel Current:}\label{met:Tunnel Current}}
%-----------------------------------------------------------------

\noindent The scanning tunneling current between the tip and the magnetic substrate can be written by using Keldysh Green's function~\cite{PhysRevLett.103.206402}. If the electron fields in the  substrate and tip are $\psi_{\rm{s}}$ and $\psi_{\rm{t}}$ respectively, the local tunneling between the substrate and  tip is then 
%------------------------------
\begin{equation}\label{seq.1}
\mathcal{H}({\bf A}) = -t\left( \psi\dg_{\rm{s}}e^{-i \frac{e}{\hbar}\int_t^s {\bf A} \cdot d\bx } \psi_{\rm{t}} + \rm{h.c.} \right),
\end{equation}
%------------------------------
where $t$ is the hopping amplitude, $\bf A$ the vector potential and we use the convention that the electron charge is $-e$ ($e>0$). Differentiating with respect to the vector potential gives the current operator 
%-----------------------------
\begin{equation}\label{seq.2}
\hat I_{t\rightarrow s} = -\left.  \frac{\delta H({\bf A})}{\delta {\bf A}}\right \vert_{{\bf A}=0} = -\frac{i e t}{\hbar}\left( \psi\dg_s \psi_t-\psi\dg_t\psi_s \right)
\end{equation}
%-----------------------------
so that the expectation of the tunneling current is given by
%-----------------------------
\begin{equation}\label{seq.3}
\langle \hat I_{t\rightarrow s} \rangle  
=
-
{\rm Re}\left[ 
\frac{i e t }{\hbar} \langle\left\{\psi_s, \psi\dg_t\right\} \rangle 
\right] = \frac{et}{\hbar}\left.{\rm Re}{\cal G}_{st}^K(t)\right\vert_{t=0}
\end{equation}
%-----------------------------
where ${\cal G}^K_{st}(t)= - i \langle \{ \psi_s(t),\psi\dg_t(0)\}\rangle $ is the Keldysh Green's function for propagation between tip and substrate. 
Following Ref.~\cite{PhysRevLett.103.206402,PhysRevLett.104.187202},  we now write the equal time Keldysh Green's function as 
%------------------------------
\begin{equation}\label{seq.4}
I(V) 
= 
\frac{et}{\hbar} {\rm{Re}} \int \frac{d\omega}{2\pi} {\rm{Tr}}\; {\cal G}^K_{\rm{st}}(\omega)
=
-\frac{e|t|^2}{\hbar} {\rm{Re}} \int \frac{d\omega}{2\pi} {\rm{Tr}}
\left[
{\cal G}^R_{\rm{s}} (\omega_-) {\cal G}^K_{\rm{t}}(\omega_+) + {\cal G}^K_{\rm{s}}(\omega_-) {\cal G}^A_{\rm{t}}(\omega_+) \right],
\end{equation}
%------------------------------
where $\omega_\pm = \omega \pm \tfrac{eV}{2}$ while ${\cal G}^R_{\rm{s(t)}}(\omega)$, ${\cal G}^A_{\rm{s(t)}}(\omega)$ and ${\cal G}^{\rm{K}}_{\rm{s(t)}}(\omega)$ and the retarded, advanced and  Keldysh Green's function for the substrate (tip) respectively, and we have used the relation ${\cal G}^K_{\rm{st}}(\omega)= -t \left[
{\cal G}^R_{\rm{s}} (\omega_-) {\cal G}^K_{\rm{t}}(\omega_+) + {\cal G}^K_{\rm{s}}(\omega_-) {\cal G}^A_{\rm{t}}(\omega_+) \right]$. Assuming the tip and the substrate at thermal equilibrium, we utilize the fluctuation-dissipation theorem~\cite{Kamenev2009}  
$${\cal G}^K(\omega) =  \left[ {\cal G}^R(\omega) - {\cal G}^A(\omega) \right] \tanh \tfrac{\beta \omega}{2}.$$
Consequently, we can rewrite the tunneling current in Eq.~\eqref{seq.2} as 
%------------------------------
\begin{equation}\label{seq.5}
I (V, \bx) = \frac{2\pi e|t|^2}{\hbar}\int d\omega {\rm Tr}\left[
g_{\rm{s}} (\omega-eV,\bx ) g_{\rm{t}} (\omega)\right]
\big[f(\omega-eV)-f(\omega)\big],
\end{equation}
%------------------------------
where $g_{\rm{s,t}} (\omega,\bx ) = \frac{1}{\pi}{\rm Im} {\cal G}^A_{\rm{s,t}} (\omega,\bx ) = \frac{1}{\pi}{\rm Im}{\cal G}_{\rm{s,t}} (\omega-i\delta,\bx ) $ are the spin-dependent density of states in the substrate (s) and tip (t), respectively, and $f(\omega) = 1/(e^{\beta\omega}+1) $ is the Fermi-Dirac function.  

If we perform a spatial Fourier transform of the tunneling current,  then 
%------------------------------
\begin{equation}\label{seq.6}
I (V, {\bf q}) 
= 
\frac{2\pi e|t|^2}{\hbar}\int d\omega {\rm Tr}\left[
g_{\rm{s}} (\omega-eV, {\bf q}) g_{\rm{t}} (\omega)\right]
\big[f(\omega-eV)-f(\omega)\big],
\end{equation}
%------------------------------
where $I(V,{\bf q}) 
= 
\frac{1}{N} \sum_{\bx_j} I(V, \bx_j) e^{i\bx_j \cdot \bq} 
$ and $g_{\rm s}(\omega, {\bq}) = \frac{1}{N} \sum_{\bx_j} g_{\rm s} (V, \bx_j) e^{i\bx_j \cdot \bq} $ are Fourier transforms of the current and the substrate density of states. 
Now, the AFM substrate develops a commensurate striped structure on the iron lattice with the reciprocal lattice wavevector  $\vq = (\pi,0)$.   The Fourier transform for the tunneling current at this wavevector can be decomposed in terms of the current at the up and down spin sites, 
%------------------------------
\begin{equation}\label{seq.7}
I(V, \vq) = \overline{I(V, \bx_\uparrow) - I(V, \bx_\downarrow)},
\end{equation}
%------------------------------
and in this way, the staggered tunneling current reflects the underlying magnetization of the substrate. \\

%-----------------------------------------------------------------------------------------------------
\subsection*{\textbf{Extraction of magnetic tunneling signal from data}\label{met:Spectrum_Analysis}}
%-----------------------------------------------------------------------------------------------------

On substituting $m_t(\omega,V) = m_0+V X(\omega)$ into equation \eqref{eq.6}, we obtain
%----------------------------------
\begin{equation}\label{meq.6}
i(V) 
= \int d\omega 
\big[f(\omega-eV)-f(\omega)\big]\overbrace{(m_0+V X(\omega))}^{m_{\rm{t}}(\omega,V)} ,
\end{equation}
%----------------------------------
where $i(V)=I(V)/I_0$ is the normalized current. In other words the conductance 
%----------------------------------
\begin{equation}\label{meq.7}
{\sf G}(eV)= \left(\frac{i(V)}{eV}\right)
= m_0 +\int d\omega 
\big[f(\omega-eV)-f(\omega)\big]X(\omega),
\end{equation}
%---------------------------
so that 
%----------------------------------
\begin{eqnarray}\label{X}
m_0&=& {\sf G}(0),\cr
{\sf G}'(eV) &= &\int d\omega \left[-f'(\omega-eV)\right] X(\omega) \equiv - f' * X.
  \end{eqnarray}  
%---------------------------
In other words, the derivative of the conductance is a convolution of $X$ with the thermally broadened delta function $\delta_T(\omega - eV)=-f'(\omega-eV)$. Temperature thus acts as a low-pass filter  so that the measured  $ X_T(eV)={\sf G}'(eV)$ is the intrinsic spectrum $X(eV)$ with fluctuations on a scales $k_BT$ and below removed. At our base temperature of 3K,  the approximation of replacing the derivative of the Fermi-Dirac function by a delta function thus coarse-grains the spectrum on scales of 3K$\sim$ 0.3meV.  Since there is no detail on these scales in the measured $I(eV,{\bf Q})$,  the approximation $X(eV) \approx X_T(eV)$ is a good  way to extract the axionic density of states. From the extrapolated $m_0=\lim_{x\rightarrow 0} {\sf G}(x)$ we can also place bounds on the static magnetization $m_0$. 

%--------------------------------------------------------------------------------------------------------------------------
\subsection*{\textbf{Absence of spin-momentum locking in the tip density of states:}\label{met:Absence of spin-momentum}}
%--------------------------------------------------------------------------------------------------------------------------

\noindent To demonstrate that the  sum over  momenta in the local density of states for the tunneling tip removes the spin-momentum locking we adopt a simple \blue{Dirac} Hamiltonian model for the topological surface states at the tip as follows:
%------------------------------
\begin{equation}\label{seq.12}
{\cal H} = v_{\rm F} \bm{\sigma} \cdot \vk,
\end{equation}
%------------------------------
where $\bm{\sigma} = (\sigma_x, \sigma_y)$ are the components of the Pauli matrices denoting the spin degrees of freedom, and $\vk = (k_x,k_y)$ are the two-dimensional momenta on the surface of SmB$_6$.
Here, $v_{\rm F}$ is the Dirac velocity of the topological surface states.  

The momentum resolved Green's function is given by:
%------------------------------
\begin{equation}\label{seq.13}
{\cal G}(\vk,\omega) = \frac{1}{\omega - v_{\rm F} \bm{\sigma} \cdot \vk + i\delta},
\end{equation}
%------------------------------
where $\delta$ is a small positive number. In momentum space, there is a clear spin-momentum locking. 

However, the local density of states relevant to STM tunneling with single-atom resolution, is obtained by summing the Green's function over  momentum space: 
%------------------------------
\begin{equation}\label{seq.14}
{\cal G}(\omega) 
= 
\sum_{\vk} \frac{1}{\omega - v_{\rm F} \bm{\sigma} \cdot \vk + i \delta}
\equiv
\sum_{\vk} \left.\frac{z + v_{\rm F} \bm{\sigma} \cdot \vk}{z^2 - v^2_{\rm F} |\vk|^2}\right\vert_{z \to \omega + i \delta}
=
\int _0^{\Lambda} \frac{k dk}{2\pi}\left.\frac{z}{z^2 - v^2_{\rm F} |\vk|^2}\right\vert_{z \to \omega + i \delta},
\end{equation}
%------------------------------
where we have introduced an ultraviolet cut-off $\Lambda$. This yields the local Green's function as:
%------------------------------
\begin{equation}\label{seq.15}
{\cal G}(\omega) 
=
-\frac{z}{4\pi v^2_{\rm F}} \left.\ln \big[ \frac{\Lambda^2 - z^2}{z^2} \big]\right\vert_{z \to \omega + i \delta}.
\end{equation}
leading to a spin-independent density of states $A(\omega) = -\frac{1}{\pi}Im[G(\omega + i \delta )] = \frac{\omega}{4 \pi v^2_{\rm F}}$.
%------------------------------
In this way, the summation over momentum eliminates all spin-dependence in the local Green's function. Qualitatively, this is a simple consequence of Heisenberg's uncertainty principle: single-atom resolution implies an uncertainty in position $\Delta x \ll a$, the unit cell size, which in turn implies an uncertainty in momentum $\Delta p \gg \frac{\hbar}{a} $ that is greater than the Brillouin zone size, eliminating all memory of momentum-spin locking. 

%----------------------------------------------------------------------------------------
\subsection*{\textbf{Weak-tunneling limit and nanowire-tip geometry:}\label{met:Facets}}
%----------------------------------------------------------------------------------------

\noindent {\color{black} The argument above shows that a locally resolved STM measurement does not directly retain the spin-momentum locking of a momentum-resolved surface Dirac cone. We now generalize this argument to incorporate the
multi-faceted nanowire-tip geometry. When a tunneling tip at meV voltages is lowered towards a substrate, the tiny pA currents that develop correspond to the opening up of the minimal-action WKB tunneling path. To exponential accuracy, the order of magnitude of the tunneling conductance is given by the Bardeen-Landauer relation
%----------------------------------
\begin{equation}\label{tunnelex}
G_0 \sim  \frac{e^2}{h} \times
\exp \left[
- 2\overbrace{
\frac{1}{\hbar} \int _t^s \sqrt{2m V({\bf x}) } dx
}^{S_E/\hbar}
\right]
\ll \frac{e^2}{h}.
\end{equation}
%----------------------------------
Here $S_E$ is the Euclidean action for the passage from the tip to the substrate~\cite{PhysRevB.31.805,Coleman_1977}. When the measured conductance is small compared with $e^2/h$, the tunneling is weak and the current can be treated to leading quadratic order in the tunneling amplitude. In this limit, the leading-order current is determined by a convolution of equilibrium tip and substrate retarded Green's functions with Fermi functions~\cite{Meir_Wingreen_1992}. In our experiments, $G=I/V\sim 10^{-4}e^2/h$, corresponding to a representative conductance of $60{\rm pA}/30{\rm mV}\sim 2\times 10^{-9}\Omega^{-1}$, compared with $e^2/h=3.87\times 10^{-5}\Omega^{-1}$.

In the weak-tunneling limit, the STM current is controlled by the local Green's function of the apex and substrate orbitals at the end of the WKB tunneling path,
%----------------------------------
\begin{equation}\label{tuncur}
I(V,\bx) \propto
G_0
\int d\omega
\biggl[
f(\omega-eV)-f(\omega)
\biggr]
{\rm Tr}\left[
{\cal G}''_A(\omega-eV)
{\cal G}''_s(\bx,\omega)
\right],
\end{equation}
%----------------------------------
where ${\cal G}_A(\omega)$ and ${\cal G}_s(\bx,\omega)$ are the equilibrium retarded Green's functions of the apex of the tip and the substrate at position $\bx$, respectively, and ${\cal G}''={\rm Im}{\cal G}$ denotes the imaginary part. (See equation \eqref{seq.5}). When the substrate is magnetic, the substrate Green's function contains a magnetic component,
%----------------------------------
\begin{equation}
{\cal G}''_{\rm s}(\bx,\omega)
=
g_{\rm s}^0(\bx,\omega)\underline{1}
+
{\bf m}_{\rm s}(\bx,\omega)\cdot{\bm \sigma}.
\end{equation}
%----------------------------------
A spin-sensitive tunneling signal can then occur only if the trace in \eqref{tuncur} is finite, which requires that the apex Green's function also contains a magnetic component,
%----------------------------------
\begin{equation}
{\cal G}''_{\rm A}(\omega)
=
g_{\rm A}^0(\omega)\underline{1}
+
{\bf g}_{\rm A}(\omega)\cdot{\bm \sigma}.
\end{equation}
%----------------------------------
Thus, in the weak-tunneling limit, spin sensitivity requires a finite magnetic part of the local apex spectral function, ${\bf g}_{\rm A}(\omega)\neq 0$.

We may model the nanowire as a collection of surface facets which hybridize with the apex orbital. Let us write the Hamiltonian of the nanowire facets, apex orbital, and substrate as
%--------------------------------------------
\begin{equation}\label{eq:facet_apex_total}
{\cal H}
=
\left
[
{\cal H}_{\rm F}
+
{\cal H}_{\rm F-A}
+
{\cal H}_{\rm A}\right]
+
{\cal H}_{\rm A-S}
+
{\cal H}_{\rm S}.
\end{equation}
%--------------------------------------------
Here ${\cal H}_{\rm F}$ describes the surface states on the different nanowire facets, ${\cal H}_{\rm F-A}$ couples the facets to the apex, ${\cal H}_{\rm A}$ describes the apex orbital connected to the tunneling path, ${\cal H}_{\rm A-S}$ describes the actual STM tunneling process between the apex and the substrate, and ${\cal H}_{\rm S}$ describes the magnetically polarized Fe$_{1+x}$Te substrate. We now show that if the tip Hamiltonian - the term inside the square brackets- is time-reversal invariant, the spin-polarized component of the tunneling current vanishes.

The topological facet states may be modeled as
%--------------------------------------------
\begin{equation}\label{eq:facet_hamiltonian}
{\cal H}_{\rm F}
=
\sum_{a,\lambda,\mathbf{k}}
c^\dagger_{a\lambda\mathbf{k}}
h_{a\lambda}(\mathbf{k})
c_{a\lambda\mathbf{k}},
\end{equation}
%--------------------------------------------
with
%--------------------------------------------------
\begin{equation}\label{eq:facet_dirac_hamiltonian}
h_{a\lambda}(\mathbf{k})
=
\epsilon_{a\lambda}(\mathbf{k})\mathbb{I}_2
+
v_{a\lambda}{\bm \sigma}\cdot
\left(\mathbf{k}\times\hat{\mathbf n}_a\right).
\end{equation}
%--------------------------------------------------
Here $c^\dagger_{a\lambda\mathbf{k}}$ creates an electron on nanowire facet $a$, where $\lambda$ labels the surface Dirac cone or pocket on that facet,  $\hat{\mathbf n}_a$ is the facet normal while ${\bf k}$ is the momentum of the gapless surface state. We have suppressed the spin indices on the creation and annihilation operators and the Hamiltonian $h_{a\lambda}(\mathbf{k})$. The apex orbital is coupled to these facet states by the facet-apex coupling,
%--------------------------------------------------
\begin{equation}\label{eq:facet_apex_coupling}
{\cal H}_{\rm F-A}
=
\sum_{a,\lambda,\mathbf{k}}
\left[
b^\dagger W_{a\lambda\mathbf{k}}c_{a\lambda\mathbf{k}}
+
c^\dagger_{a\lambda\mathbf{k}}W\dg_{a\lambda\mathbf{k}}b
\right],
\end{equation}
%--------------------------------------------------
where $b\dg= (b\dg_{\uparrow}, b\dg_{\downarrow})$ creates an electron in the apex orbital.  The tunneling Hamiltonian is
%--------------------------------------------------
\begin{equation}\label{eq:apex_substrate_tunneling}
{\cal H}_{\rm A-S}
=
d^\dagger(\mathbf r_0)Tb
+
b^\dagger T^\dagger d(\mathbf r_0),
\end{equation}
%--------------------------------------------------
where the magnitude of the tunneling matrix element has the WKB form shown in Eq.~\eqref{tunnelex}.

Integrating out the facet states gives the apex Green's function
%--------------------------------------------------
\begin{equation}\label{eq:apex_green_function}
{\cal G}_{\rm A}(\omega)
=
\frac{1}
{
\omega-\epsilon_0+i0^+
-
\Sigma_{\rm A}(\omega)
},
\end{equation}
%--------------------------------------------------
where we use the matrix notation $1/A\equiv A^{-1}$ and
%--------------------------------------------------
\begin{equation}\label{eq:facet_self_energy}
\Sigma_{\rm A}(\omega)
=
\sum_{a,\lambda,\mathbf{k}}
W_{a\lambda\mathbf{k}}
\frac{1}{\omega+i0^+-h_{a\lambda}(\mathbf{k})}
W\dg_{a\lambda\mathbf{k}}.
\end{equation}
%--------------------------------------------------
This self-energy contains a sum over contributions from the Dirac cones $\lambda$ on each facet $a$. We split it into diagonal and magnetic components,
%-------------------
\begin{equation}
\Sigma_{\rm A}(\omega)
=
\Sigma^0_{\rm A}(\omega)\underline{1}
+
{\bm \Sigma}_{\rm A}(\omega)\cdot{\bm \sigma}.
\end{equation}
%-------------------
We now show that the magnetic term ${\bm \Sigma}_{\rm A}(\omega)$ is odd under time reversal, so that if the tip preserves time-reversal symmetry, there is no magnetic component to the apex Green's function.

First, we rewrite the self-energy in the form
%-------------------
\begin{equation}
\Sigma_{\rm A}(\omega)
=
\sum_{a,\lambda,\mathbf{k}}
W_{a\lambda\mathbf{k}}
\frac{
\omega-\epsilon_{a\lambda}
+
v_{a\lambda}
(\mathbf{k}\times\hat{\mathbf n}_a)\cdot{\bm \sigma}
}
{
[\omega-\epsilon_{a\lambda}+i0^+]^2
-
v_{a\lambda}^2 k^2
}
W\dg_{a\lambda\mathbf{k}}.
\end{equation}
%-------------------
We can extract the spin-diagonal and spin-dependent components of the self-energy from
%-------------------
\begin{eqnarray}
\Sigma^0_{\rm A}(\omega)
&=&
\frac{1}{2}{\rm Tr}[\Sigma_{\rm A}(\omega)]
\nonumber \\
&=&
\frac{1}{2}
\sum_{a,\lambda,\mathbf{k}}
{\rm Tr}
\left[
\frac{
\omega-\epsilon_{a\lambda}
+
v_{a\lambda}
(\mathbf{k}\times\hat{\mathbf n}_a)\cdot{\bm \sigma}
}
{
[\omega-\epsilon_{a\lambda}+i0^+]^2
-
v_{a\lambda}^2 k^2
}
W\dg_{a\lambda\mathbf{k}}W_{a\lambda\mathbf{k}}
\right],
\cr
{\bm \Sigma}_{\rm A}(\omega)
&=&
\frac{1}{2}{\rm Tr}[\Sigma_{\rm A}(\omega){\bm \sigma}]
\nonumber \\ \label{spind}
&=&
\frac{1}{2}
\sum_{a,\lambda,\mathbf{k}}
{\rm Tr}
\left[
\frac{
\omega-\epsilon_{a\lambda}
+
v_{a\lambda}
(\mathbf{k}\times\hat{\mathbf n}_a)\cdot{\bm \sigma}
}
{
[\omega-\epsilon_{a\lambda}+i0^+]^2
-
v_{a\lambda}^2 k^2
}
W\dg_{a\lambda\mathbf{k}}{\bm \sigma}W_{a\lambda\mathbf{k}}
\right].
\end{eqnarray}
%-------------------
Now the time-reversal symmetry operation is given by $\Theta=-i\sigma_2K,$ where $K$ is the complex conjugation operator. Under this transformation,
%-------------------
\begin{equation}
W_{a\lambda\mathbf{k}}
\rightarrow
\Theta W_{a\lambda\mathbf{k}}\Theta^{-1}
=
-i\sigma_2 W^*_{a\lambda-\mathbf{k}} i\sigma_2,
\end{equation}
%-------------------
while the facet Hamiltonian is time-reversal invariant,
%-------------------
\begin{equation}
\Theta
\left[
(\mathbf{k}\times\hat{\mathbf n}_a)\cdot{\bm \sigma}
\right]
\Theta^{-1}
=
\left[
(-\mathbf{k})\times\hat{\mathbf n}_a
\right]\cdot(-{\bm \sigma})
=
(\mathbf{k}\times\hat{\mathbf n}_a)\cdot{\bm \sigma}.
\end{equation}
%-------------------
These relationships hold in the presence of spin-orbit coupling, so that under time reversal,
%--------------------------------
\begin{eqnarray}\label{trev}
\Theta{\bm \Sigma}_{\rm A}(\omega)\Theta^{-1}
&=&
\frac{1}{2} \sum_{a,\lambda,\mathbf{k}} {\rm Tr} \left[ \frac{\omega-\epsilon_{a\lambda} + v_{a\lambda} (\vk \times\hat{\mathbf n}_a)\cdot{\bm \sigma}}
{[\omega-\epsilon_{a\lambda}+i0^+]^2 - v_{a\lambda}^2 k^2}  \bigl[(-i\sigma_2)W^T_{a\lambda-\mathbf{k}}
\overbrace{(i\sigma_2){\bm \sigma}(-i\sigma_2)}^{-{\bm \sigma}^T}
W^*_{a\lambda-\mathbf{k}}(i\sigma_2) \bigr] \right] \cr
&=&
-\frac{1}{2} \sum_{a,\lambda,\mathbf{k}} {\rm Tr} \left[ \frac{ \omega-\epsilon_{a\lambda} - v_{a\lambda}(\vk\times\hat{\mathbf n}_a)\cdot{\bm \sigma}}
{[\omega-\epsilon_{a\lambda}+i0^+]^2 - v_{a\lambda}^2 k^2}  \bigl[ -i\sigma_2 (W^T_{a\lambda\mathbf{k}}{\bm \sigma}^T W^*_{a\lambda\mathbf{k}}) i\sigma_2 \bigr]
\right] \cr
&=&
-\frac{1}{2}
\sum_{a,\lambda,\mathbf{k}}
{\rm Tr}
\left[
\frac{
\omega-\epsilon_{a\lambda}
+
v_{a\lambda}
(\mathbf{k}\times\hat{\mathbf n}_a)\cdot{\bm \sigma}^T
}
{
[\omega-\epsilon_{a\lambda}+i0^+]^2
-
v_{a\lambda}^2 k^2
}
(W^T_{a\lambda\mathbf{k}}{\bm \sigma}^T W^*_{a\lambda\mathbf{k}})
\right].
\end{eqnarray}
%--------------------------------
On the first line we have used the fact that the facet Hamiltonian is invariant under time reversal; on the second line we have interchanged ${\bf k}\rightarrow-{\bf k}$ inside the momentum sum and used $\sigma_2{\bm \sigma}\sigma_2=-{\bm \sigma}^T$. Finally, taking the transpose of the argument inside the trace and using ${\rm Tr}A={\rm Tr}A^T$, the whole expression returns to its original form with an overall minus sign,
%--------------------------------
\begin{eqnarray}\label{id1}
\Theta{\bm \Sigma}_{\rm A}(\omega)\Theta^{-1}
&=&
-\frac{1}{2}
\sum_{a,\lambda,\mathbf{k}}
{\rm Tr}
\left[
\frac{
\omega-\epsilon_{a\lambda}
+
v_{a\lambda}
(\mathbf{k}\times\hat{\mathbf n}_a)\cdot{\bm \sigma}^T
}
{
[\omega-\epsilon_{a\lambda}+i0^+]^2
-
v_{a\lambda}^2 k^2
}
(W^T_{a\lambda\mathbf{k}}{\bm \sigma}^T W^*_{a\lambda\mathbf{k}})
\right]
\cr
&=&
-\frac{1}{2}
\sum_{a,\lambda,\mathbf{k}}
{\rm Tr}
\left[
\frac{
\omega-\epsilon_{a\lambda}
+
v_{a\lambda}
(\mathbf{k}\times\hat{\mathbf n}_a)\cdot{\bm \sigma}
}
{
[\omega-\epsilon_{a\lambda}+i0^+]^2
-
v_{a\lambda}^2 k^2
}
W\dg_{a\lambda\mathbf{k}}{\bm \sigma}W_{a\lambda\mathbf{k}}
\right]
\cr
&=&
-{\bm \Sigma}_{\rm A}(\omega),
\end{eqnarray}
%--------------------------------
establishing that the magnetic component of the self-energy is odd under time reversal.

However, if the matrix elements linking the facets to the tip are time-reversal invariant,
%-----------------------
\begin{equation}
W_{a\lambda\mathbf{k}}
=
\Theta W_{a\lambda\mathbf{k}}\Theta^{-1}
=
-i\sigma_2 W^*_{a\lambda-\mathbf{k}}i\sigma_2,
\end{equation}
%-----------------------
and similarly $ W\dg_{a\lambda\mathbf{k}} = -i\sigma_2 W^T_{a\lambda-\mathbf{k}}i\sigma_2$, so that the first line of Eq.~\eqref{trev} can be rewritten 
%----------------------------
\begin{eqnarray}\label{id2}
\Theta{\bm \Sigma}_{\rm A}(\omega)\Theta^{-1}
&=&
\frac{1}{2}
\sum_{a,\lambda,\mathbf{k}}
{\rm Tr}
\left[
\frac{
\omega-\epsilon_{a\lambda}
+
v_{a\lambda}
(\mathbf{k}\times\hat{\mathbf n}_a)\cdot{\bm \sigma}
}
{
[\omega-\epsilon_{a\lambda}+i0^+]^2
-
v_{a\lambda}^2 k^2
}
\biggl[
\overbrace{
(-i\sigma_2)W^T_{a\lambda-\mathbf{k}}(i\sigma_2)
}^{W\dg_{a\lambda\mathbf{k}}}
{\bm \sigma}
\overbrace{
(-i\sigma_2)W^*_{a\lambda-\mathbf{k}}(i\sigma_2)
}^{W_{a\lambda\mathbf{k}}}
\biggr]
\right]\cr
&=& {\bm \Sigma}_{\rm A}(\omega).
\end{eqnarray}
%----------------------------
We can only reconcile Eq.~\eqref{id1} and Eq.~\eqref{id2} if ${\bm \Sigma}_{\rm A}(\omega) = - {\bm \Sigma}_{\rm A}(\omega)=0$. Therefore, despite the spin-momentum locking of the nanowire surface facets, the magnetic component of the facet-induced self-energy vanishes at all frequencies when the nanowire tunneling tip is time-reversal invariant. In short, spin-momentum-locked surface states and their time-reversal-invariant coupling to the tunneling apex cannot create a spin-polarized STM current in the weak-tunneling limit. A finite magnetic tunneling contrast therefore requires a time-reversal-odd local spectral component of the SmB$_6$ tip, consistent with the voltage-induced magnetic response analyzed in the main text. } 

%-----------------------------------------------------------
\subsection*{\textbf{Quantifying the Strength of the Moment:}\label{met:Moment}}
%-----------------------------------------------------------

\noindent To estimate  the tip magnetization we adopt a simplified model for the tip and substrate.  We assume that the low energy  density of states in the substrate contains two components - a non-magnetic uniform component  and a magnetic component at $\vq = (\pi,0)$.  If we assume $ g_{\rm t}(\omega) \approx \rho_{\rm t}(\omega) [1 + {\bf m}_{\rm t}\cdot {\bm{\sigma}}]$ and $ g_{\rm s}(\omega, \bx) \approx \rho_{\rm s}(\omega) [1 + {\bf m}_{\rm s}\cdot {\bm{\sigma}}e^{i\vq \cdot \bx }]$, then the density of states factorizes out in the ratio between the tunnel  current at the staggered and uniform  $\bq$ vector
%------------------------------
\begin{equation}\label{seq.9}
\frac{I(V,\vq)}{I(V, {\bf q}=0)}
= 
{\bf m}_{\rm s} \cdot {\bf m}_{\rm t}
\end{equation}
%------------------------------
If we now compare this ratio for a magnetic Cr tip and an \smb tip, it follows that the ratio of the \smb and Cr moment is given by 
%------------------------------
\begin{equation}\label{seq.10}
\frac{I_{\smb}(V, \vq)}{I_{\smb}(V, {\bf q}=0)}
\bigg/
\frac{I_{\rm Cr}(V, \vq)}{I_{\rm Cr}(V, {\bf q}=0)} 
= 
\frac{\rm m_{\rm Sm}}{\rm m_{\rm Cr}}.
\end{equation}
%------------------------------
In practice, the  large noise background in the Fourier transformed signal at $\bq =0$ means we must replace the signal at $\bq =0$ by the signal at an alternative reference wave-vector.  Fortunately, the structure of Fe$_{1+x}$Te provides us with an alternate comparison point in momentum space. In a layer of Fe$_{1+x}$Te, the iron atoms are located in a square array at locations ${\bx}_{\rm Fe} = a(l,n,0)$ $(l,n \in \mathbb{Z})$, while  the Te atoms form a checker-board pattern, alternating their position above or below the centers of the iron squares, ${\bx}_{\rm Te} =  (a (l+\frac{1}{2}),a(n+\frac{1}{2}),\pm b)$. The tunneling density of states is  unpolarized at the tellurium atoms, so that we can use the Fourier transformed signal at the set-point ${\bf Q}_{\rm Te} = (\pi,\pi)$, i.e.,
%------------------------------
\begin{equation}\label{seq.11}
\frac{I_{\smb}(V, \vq)}{I_{\smb}(V, \vq_{\rm Te})}
\bigg/
\frac{I_{\rm Cr}(V, \vq)}{I_{\rm Cr}(V, \vq_{\rm Te})} 
= 
\frac{\rm m_{\rm Sm}}{\rm m_{\rm Cr}}.
\end{equation} 
%------------------------------

%-------------------------------------------------------------
\subsection*{\textbf{Axial Symmetry of the Dirac Equation:}\label{met:Axial Dirac}}
%-------------------------------------------------------------

\noindent \blue{In this section, we outline the underlying mathematics behind a Bloch domain wall on the surface of a topological insulator. The topological quantum critical point that separates a topological and conventional insulator is described by a Dirac metal, and the physics and topology of the transition from a conventional to a topological insulator may be modeled in terms of a continuum Dirac equation}
%--------------------------------
\begin{equation}
\blue{{\cal H} = \int d^3x\, \psi\dg(x) \left[ -i\bm{\alpha}\cdot\bm{\nabla} +\beta m_e \right] \psi(x).}
\end{equation}
%--------------------------------
\blue{Here $\alpha^i=\gamma^0\gamma^i$ ($i=1,2,3$) and $\beta\equiv\gamma^0$ are the Dirac matrices, while \blue{the effective Dirac mass} $m_e$ plays the role of a \blue{band gap}. The phase with $m_e>0$ represents a conventional insulator, while $m_e<0$ describes a band-inverted topological insulator.

The fifth Dirac matrix $\gamma_5=\gamma_5\dg=i\gamma^0\gamma^1\gamma^2\gamma^3$ plays an important role at a topological phase transition. The associated axial charge $Q_5=\int d^3x\,\psi\dg(x)\gamma_5\psi(x)$ commutes with the Dirac Hamiltonian when $m_e=0$ and is \blue{conserved}. $Q_5$ is the generator of \blue{a global $U(1)$ axial rotation}, $\psi(x)\rightarrow e^{i\frac{\theta}{2}\gamma_5}\psi(x)$, which leaves the action and Hamiltonian unchanged for $m_e=0$. For $m_e\neq0$, the axial rotation modifies the Hamiltonian}
%--------------------------------
\begin{equation}
\blue{{\cal H} \rightarrow {\cal H}(\theta) =
\int d^3x\, \psi\dg(x) \left[ -i\bm{\alpha}\cdot\bm{\nabla} +\beta e^{i\theta\gamma_5}m_e \right] \psi(x).}
\end{equation}
%--------------------------------
\blue{In other words, the Dirac mass term transforms according to $ m_e\rightarrow e^{i\theta\gamma_5}m_e$, where}
%--------------------------------
\begin{equation}\label{complexmass}
\blue{ e^{i\theta\gamma_5}m_e = m_e\cos\theta +i\gamma_5m_e\sin\theta \equiv m+i\gamma_5\Delta.}
\end{equation}
%--------------------------------
\blue{In the condensed-matter setting, we may therefore introduce the complex scalar order parameter ${\cal M}=m+i\Delta=|{\cal M}|e^{i\theta}$, where $(m,\Delta)=(m_e\cos\theta,m_e\sin\theta)$. A non-zero $\Delta$ corresponds to an axial mass component and breaks the continuous $U(1)$ axial symmetry.

Now suppose that $\theta(x)$ \blue{varies slowly in space and time}. We can then formally construct a Ginzburg--Landau theory for the axion angle. Integrating out the massive fermions and expanding the resulting fermion determinant at low energies gives~\cite{Li2010}}
%--------------------------------
\begin{equation}
\blue{
\begin{aligned}
S_{\rm eff}[\theta,A]
&=
-i\,{\rm Tr}\ln
\left[
i\slashed{\partial}
-e\slashed{A}
-m_e e^{i\theta(x)\gamma_5}
\right]
\\
&\simeq
S_{\rm eff}[0,A]
+
\int d^4x
\left[
\frac{\rho}{2}
(\partial_\mu\theta)(\partial^\mu\theta)
+
\frac{e^2}{2\pi h}\,
\theta(x)\,
{\mathbf E}\cdot{\mathbf B}
\right]
+\cdots .
\end{aligned}
}
\end{equation}
%--------------------------------
\blue{The fermion determinant must be carefully regularized, and the displayed expression is the leading term in a low-energy derivative
expansion. In a regularized calculation, the phase stiffness $\rho$ of the axionic field scales with $|{\cal M}|^2=m^2+\Delta^2$, up to non-universal cutoff-dependent factors, and therefore requires a finite ultraviolet cutoff. The additional term in the action is known as the axial anomaly, a topological correction to the action that appears in the presence of an
electromagnetic field and is associated with the non-trivial Jacobian of the axial rotation~\cite{Fujikawa:1979ay,Fujikawa:1980eg}. It is this term that is responsible for the magnetoelectric properties.}

\blue{At the continuum level, the Gordon decomposition separates the microscopic electromagnetic current into particle-current and spin-magnetization-current contributions~\cite{Gordon1928}. The total magnetization induced by the axion response may therefore be written as ${\mathbf M}_{\theta}={\mathbf M}_{\rm orb}+{\mathbf M}_{\rm s}$, while the anomaly fixes only their sum, ${\mathbf M}_{\theta}=(\theta e^2/2\pi h){\mathbf E}$. Their relative orbital and spin weights depend on the microscopic electronic structure of the material.}

\blue{We can model the surface of a topological insulator as a region where the mass changes sign, forming a domain wall. If the effective Dirac mass is generated by interactions, we may introduce a Gross--Neveu interaction of the form}
%--------------------------------
\begin{equation}
\blue{
S_I
=
-\frac{1}{2}
\int d^4x
\left[
g(\bar{\psi}\psi)^2
-
g_5(\bar{\psi}\gamma_5\psi)^2
\right],
\qquad
\frac{1}{g_5}
=
\frac{1}{g}
+
\frac{1}{\lambda}}.
\end{equation}
%--------------------------------
\blue{Here $\lambda$ parametrizes the additional anisotropy in the $\gamma_5$ channel. Carrying out a Hubbard--Stratonovich transformation gives
%--------------------------------
\begin{equation}
S_I
=
\int d^4x
\left[
\bar{\psi}
\left(
m(x)+i\gamma_5\Delta(x)
\right)
\psi
+
\frac{m(x)^2+\Delta(x)^2}{2g}
+
\frac{\Delta(x)^2}{2\lambda}
\right].
\end{equation}
%--------------------------------
Integrating out the fermions and carrying out a derivative and order-parameter expansion generates gradient and potential terms for $m$ and $\Delta$. After rescaling the fields and coordinates, this motivates the phenomenological free-energy functional}
%--------------------------------
\begin{equation}
\blue{
F = \frac12\int dx
\left\{
|\partial_x{\cal M}|^2
+\left(|{\cal M}|^2-1\right)^2
+\frac{\Delta(x)^2}{\lambda}
\right\}.
}
\end{equation}
%--------------------------------
\blue{This free-energy functional supports domain-wall configurations in which the scalar mass $m(x)$ changes sign, while the axial component $\Delta(x)$ may become nonzero inside the wall, allowing the complex mass ${\cal M}(x)=m(x)+i\Delta(x)$ to rotate continuously between the two bulk vacua. 
To see this explicitly, the Lagrangian equation for the order parameter is given by
\begin{equation}
    \frac{\partial F}{\partial {\cal M}^*}= \frac{1}{2}\left[  - \partial_x ^2{\cal M} + 2\left(|{\cal M}|^2-1\right){\cal M}  + i \frac{\Delta}{\lambda}\right],
\end{equation}
which has domain wall solutions of the form 
\begin{equation} \label{eq:domains} {\cal M}(x)= \left\{
\begin{array}{ll}
\tanh \left({x}\right) ,    & (\lambda <1: \hbox{Ising}) \cr
 \tanh \left(\frac{x}{\sqrt{\lambda}}\right) \pm i \sqrt{\left(1 - \frac{1}{\lambda}\right)}\  \mathrm{ sech}
 \left(\frac{x}{\sqrt{\lambda}}\right) ,    & (\lambda >1: \hbox{Bloch})
\end{array}\right.
\end{equation}
leading to Bloch domain walls developing for $\lambda > 1$  as shown in Fig. \ref{fig:Fig4X}.
In the following, we outline how an analogous Bloch-type interpolation can be realized in a lattice-regularized Wilson--Dirac model.} \\

%---------------------------------------------------------------------------------------
\subsection* {\textbf{Axial Rotation as a family of Wilson-Dirac lattice Hamiltonians.} }
\label{met:Wilson Dirac}
%---------------------------------------------------------------------------------------

\noindent \blue{The continuum complex mass can be regularized on a cubic lattice as a four-band Wilson--Dirac Hamiltonian}
%--------------------------------
\begin{equation}\label{eq:Wilson_Dirac}
\blue{
{\cal H}_{\rm WD}({\bf k},\theta)
=
V\sum_{i=1}^{3}\sin k_i\,\Gamma_i
+
\left[
M_0\cos\theta
+
t\sum_{i=1}^{3}(1-\cos k_i)
\right]\Gamma_4
+
M_0\sin\theta\,\Gamma_5 .
}
\end{equation}
\blue{In the context of topology, ${\cal H}_{\rm WD}({\bf k},k_4)$\cite{Zhang2008,BernevigHughes2013} defines a four-dimensional Chern insulator. By replacing   $k_4\rightarrow\theta\in [0,2\pi]$ a $Z_2$ topological insulator is obtained as projection of a four dimensional Chern insulator.  $\theta$ replaces the axial rotation angle of the continuum Dirac equation, while $\Gamma^{1,2,3}\sim \mathbf{\alpha}$,  $\Gamma^4\sim \beta$ and $\Gamma^5\sim \beta \gamma_5$ replace the corresponding continuum matrices.}
%--------------------------------
\blue{A convenient representation of the mutually anticommuting Dirac matrices is}
%--------------------------------
\begin{equation}\label{Diracs}
\blue{
\Gamma_i=\tau_x\otimes\sigma_i
\quad (i=1,2,3),
\qquad
\Gamma_4=\tau_z\otimes\sigma_0,
\qquad
\Gamma_5=\tau_y\otimes\sigma_0,
}
\end{equation}
%--------------------------------
\blue{where the Pauli matrices $\tau_i$ and $\sigma_i$ act in orbital and spin space, respectively, 
so that
%--------------------
\begin{equation}
{\cal H}_{\rm WD}({\bf k},\theta) 
= 
\left(
\begin{matrix}
t \sum_l (1- \cos k_l) + M_0\cos\theta  & V \sin {\mathbf{k}}\cdot \bm{\sigma}  -iM_0\sin\theta\cr V\sin {\mathbf{k}}\cdot\bm{\sigma}+ iM_0\sin\theta & -t \sum_l (1- \cos k_l)  -  M_0\cos\theta
\end{matrix}
\right).
\end{equation}
%--------------------
As in the continuum Eq.~\eqref{complexmass}, the two components  $m_\theta=M_0\cos\theta,
\ 
\Delta_\theta=M_0\sin\theta,$ can be combined into a complex  mass }
%--------------------------------
$\blue{{\cal M}=m_{\theta}+i\Delta_{\theta}=M_0e^{i\theta}.}$

\blue{The energy spectrum is}
%--------------------------------
\begin{equation}
\blue{
E_{\pm}({\bf k},\theta)
=\pm \left[
V^2\sum_i\sin^2 k_i
+
\left[
m_{\theta}+
t\sum_i(1-\cos k_i)
\right]^2
+
\Delta_{\theta}^2\right]^{1/2}.
}
\end{equation}
%-------------------------------
\blue{For $0<M_0<2t$, the states at $\theta=0$ and $\theta=\pi$ represent conventional and band-inverted insulating limits, while the spectrum remains fully gapped at all intermediate values of $\theta$.} \blue{We can rewrite ${\cal H}_{\rm WD}({\bf k},\theta)$ to emphasize its connection with a topological insulator,}
%------------------------
\begin{equation}
\blue{{\cal H}_{\rm WD}({\bf k},\theta) 
= 
\left(
\begin{matrix}
\epsilon_c({\bf k},\theta) & V {\bf d}({\bf k}) \cdot \bm{\sigma}  -i\Delta_{\theta} \cr 
V{\bf d}({\bf k}) \cdot\bm{\sigma}+ i\Delta_{\theta} & \epsilon_f({\bf k},\theta)
\end{matrix}
\right),}
\end{equation}
%------------------------
\blue{This is a minimal Hamiltonian for a Kondo insulator\cite{Dzero2016} formed between conduction electrons with tight-binding dispersion $\epsilon_{c}(\mathbf{k},\theta) = [t \sum_l (1- \cos k_l) + m_{\theta}]$ and f-electrons with dispersion 
$\epsilon_{f}(\mathbf{k},\theta) = -[t \sum_l (1- \cos k_l) + m_{\theta}]$ via a  spin-orbit coupled hybridization $V({\bf k}) =  V{\bf d}({\bf k}) \cdot \bm{\sigma} $ where ${\bf d}({\bf k}) = (\sin k_x,\sin k_y, \sin k_z)$. Here we have adopted the $\bf d$ notation to delineate the topological d-vector from the momenta. The  anomalous term $i \Delta_{\theta}$ which appears for $\sin\theta \neq 0$ is momentum-independent, thus describing a {\sl local} hybridization between $c$ and $f$-electrons. This local Kondo hybridization between odd-parity $f$-electrons and even parity conduction electrons (d-electrons in SmB$_6$) breaks the inversion and time reversal symmetry, while preserving their product.}

\clearpage

%---------------------------
\noindent
\textbf{Acknowledgments} 
We thank Alexander V. Balatsky, Daniel Kaplan, Taylor Hughes, Anirvan Sengupta and  David Vanderbilt  for their insightful discussions.
 This work was supported by the Office of Basic Energy Sciences, Material Sciences and Engineering Division, U.S. Department of Energy (DOE) under Contracts  No. DE-FG02-99ER45790 (SB and PC) and  DE-FG02-84ER45118 (EM). VM acknowledges support from the US Department of Energy (DOE), Office of Science, Office of Basic Energy Sciences (BES), Materials Sciences and Engineering Division under award No. DE-SC0022101, and the Gordon and Betty Moore Foundation EPiQS initiative through grant number GBMF9465. \\

\noindent 
\textbf{Additional information} Correspondence and requests for materials should be addressed to P. Coleman (piers.coleman@rhul.ac.uk), V. Madhavan (vm1@illinois.edu) or E. J. Mele (mele@sas.upenn.edu). \\

\noindent
\textbf{Author contributions} The project was conceived by P.C. and V. M. The nanowires were grown by F.L. and L. J.  STM measurements and analysis were carried out by A. A. and V. M.  The theoretical analysis and conceptual framework was developed by S.B., P.C. and E. J. M. who contributed equally to this paper. All authors participated in discussions.\\

\noindent
\textbf{Data and materials availability:} The data that support the findings of this study are available from the corresponding authors upon reasonable request. \\

\noindent
\textbf{Competing financial interests} The authors declare no competing financial interests. \\

\clearpage

%------------------------------
\bibliography{Ref} 
\bibliographystyle{naturemag}
%------------------------------

%---------------------------------------------------------------------------------------------------------------------
% Here goes all the pertinent figures -- Fig 1 -- Explaining the main setup of the experiment, needs to be modified
%---------------------------------------------------------------------------------------------------------------------

\pagebreak
%-----------------------------------------------

\vskip 2in
\pagebreak
\vfill\eject
%-----------------------------------------------

\newpage
\renewcommand{\figurename}{Extended Figure}
\renewcommand{\thefigure}{S\arabic{figure}}
\setcounter{figure}{0}
%-----------------------------------------------
\begin{figure} [t]
\centering
\includegraphics[width=1.0\textwidth]{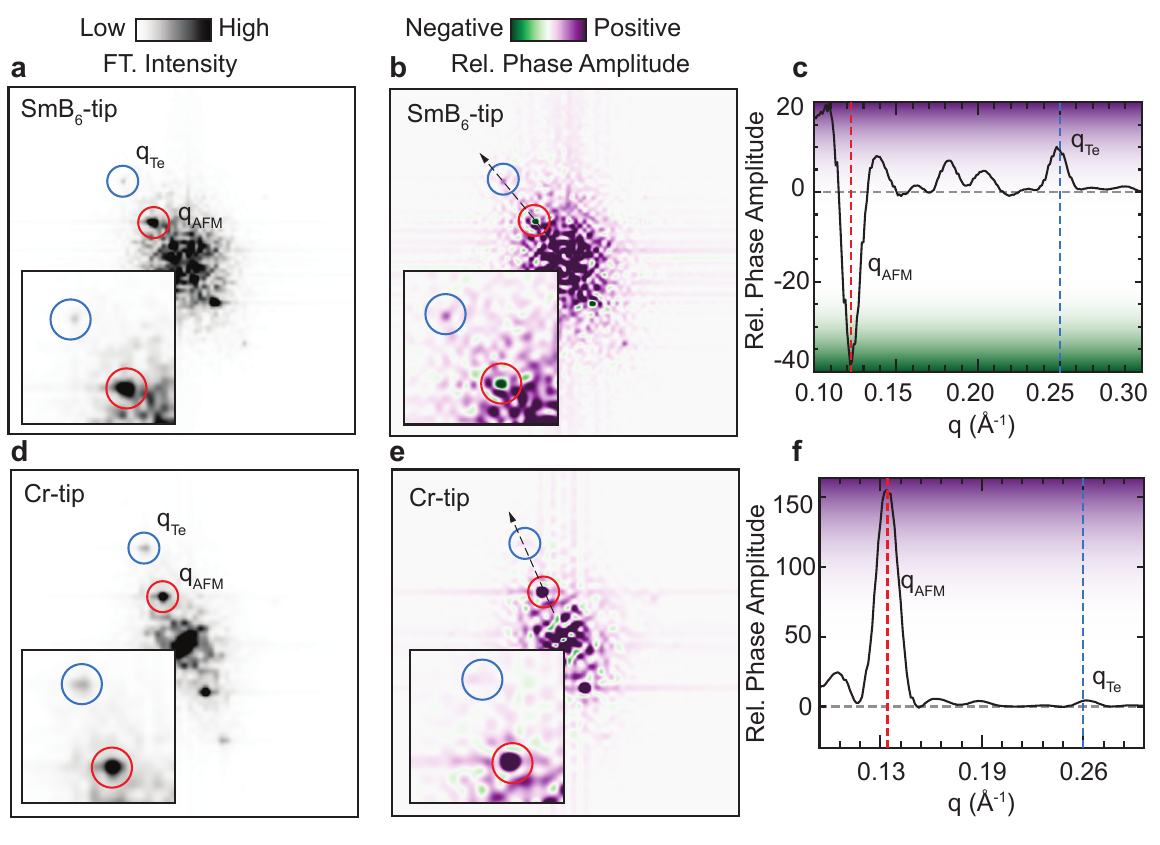}
\caption{\textbf{Spin contrast.} Spin contrast comparison of the tunneling between the SmB$_6$ ({\bf a}, {\bf b}, {\bf c}) and Cr ({\bf d}, {\bf e}, {\bf f}) tips into the anti-ferromagnetic (AFM) sample Fe$_{1+x}$Te. {\bf a}, {\bf b}, {\bf d}, {\bf e} Phase-referenced FFT amplitudes of the height-profile obtained between $\pm 30$ \blue{mV}. Relative phase amplitudes as a function of the AFM wave vector explicitly showing the contrast between SmB$_6$ ({\bf c}) and Cr ({\bf f}) tips. Reproduced from Ref.~\cite{Aishwarya2022}.}\label{fig:Fig4}
\end{figure}
%-----------------------------------------------

%\nolinenumbers

%-----------------
\end{document}